\documentclass[reprint,amsmath,amssymbGaps,nofootinbib,onecolumn,notitlepage]{revtex4-1}

\usepackage{amsmath} 
\usepackage{amssymb}
\usepackage{amsfonts}
\usepackage{amstext}
\usepackage{graphicx}
\usepackage{bm}
\usepackage{color} 
\usepackage{transparent}
\usepackage{mathtools}
\usepackage{slashed}
\usepackage{hyperref}

\makeatletter
\renewcommand{\p@subsection}{}
\renewcommand{\p@subsubsection}{}
\makeatother

\numberwithin{equation}{section}

\newcommand{\ze}{\kern 0.05em}

\begin{document}

\title{Einstein-Maxwell-scalar black holes with massive and self-interacting scalar hair}

\author{Pedro G. S. Fernandes}
 \email{p.g.s.fernandes@qmul.ac.uk}
\affiliation{School of Physics and Astronomy, Queen Mary University of London, Mile End Road, London, E1 4NS, UK}


\begin{abstract}

\par Recently, spontaneous scalarization of charged black holes has attracted a great deal of attention and motivated several studies of Einstein-Maxwell-scalar models. These studies have, however, only considered a massless and non-self-interacting scalar field. In this work a more realistic treatment of the problem is considered by studying the effects of scalar field mass and self-interacting terms on spontaneous scalarization in Einstein-Maxwell-scalar models, and on the string theory motivated dilatonic black hole. We assess the domains of existence of the black hole solutions, thermodynamic preference and radial profiles. Then we discuss on the stability of the scalarized solutions, focusing on spherical perturbations, finding all studied solutions to be stable.

\end{abstract}

\maketitle

%
 \section{Introduction}\label{S1}
%

Einstein-Maxwell-scalar (EMS) models are generically described by the action (in units with $8\pi G=c=1$)
\begin{equation}
    S = \int d ^ { 4 } x \sqrt { - g } \left[R - 2 \partial _ { \mu } \phi \partial ^ { \mu } \phi - f(\phi) F_{\mu \nu} F^{\mu \nu} - U(\phi) \right]\ ,
    \label{eq:action}
\end{equation}
that describes a real scalar field $\phi$ with a potential $U(\phi)$ minimally coupled to Einstein's gravity ($R$ is the Ricci scalar) and non-minimally coupled to Maxwell's electromagnetism ($F_{\mu \nu}=\partial_\mu A_\nu - \partial_\nu A_\mu$ is the usual Maxwell tensor) through a function $f(\phi)$.
Non-minimal couplings between the electromagnetic field and a scalar field, and the corresponding black hole (BH) solutions, have long been considered in the context of, e.g., Kaluza-Klein theory and supergravity \cite{DilatonGarfinkle,DilatonGibbons}. More recently, EMS models have been shown to allow the phenomenon of spontaneous scalarization of asymptotically flat, charged BHs \cite{MainPaper1,MainPaper2,MainPaper3,MainPaper4} (see, \textit{e.g.}, \cite{Doneva:2010ke,Stefanov:2007eq,Gubser:2005ih} for earlier discussions of charged BH scalarization in different models). Spontaneous scalarization, also considered in a BH context in extended-Scalar-Tensor-Gauss-Bonnet (eSTGB) models \cite{Macedo:2019sem,Doneva:2019vuh,Silva_2018,Doneva_2018,Antoniou_2018,Cunha:2019dwb}, is a strong gravity phase transition. It occurs when two phases (classes of solutions) co-exist, one of them becoming dynamically preferred. In EMS models, for certain choices of the coupling function $f(\phi)$, the standard electrovacuum (scalar-free) Reissner-Nordstr\"om (RN) BH solves the equations of motion, along with a new class of BHs that allow a non-trivial equilibrium scalar-field configuration (scalar-hair). For sufficiently large BH charge to mass ratio the RN BH becomes unstable against scalar perturbations and the formation of these hairy BHs is conjectured to be the endpoint of the instability \cite{MainPaper1,MainPaper2,MainPaper4}.

\par Astrophysical charged black holes are typically dismissed due to the electrical neutrality of the universe \cite{Zajacek:2019kla}, quantum discharge effects \cite{Gibbons:1975kk}, among others. There are, however, scenarios (that typically invoke \textit{new physics} arguments) where charged black holes can be of astrophysical relevance. One of such scenarios is a class of self-interacting dark matter models where the interaction is mediated by some \textit{dark} photon with no coupling to the Standard Model particles (see, \textit{e.g.}, \cite{Cardoso:2016olt, DeRujula:1989fe, Plestid:2020kdm, Ackerman:mha}). These studies of minicharged dark matter tell us that there might exist astrophysical black holes with considerably large charge to mass ratio. Such charge to mass ratio, if sufficiently large, could trigger spontaneous scalarization in EMS models, possibly leading to distinct observational signatures both in gravitational waves \cite{Abbott:2016blz} and in black hole shadows \cite{Psaltis:2018xkc}. In any case, from a theoretical point of view, the study of charged solutions gives a more complete picture of the physics of black holes and spontaneous scalarization.

\par In EMS models, studies up to now have focused on the simple case of massless and non-self-interacting scalar field \cite{MainPaper1,MainPaper2,MainPaper3,MainPaper4,HotColdBald,Myung:2018vug,Boskovic:2018lkj,EMS-Stability,Brihaye:2019kvj,Herdeiro:2019oqp,Myung:2019oua,Konoplya:2019goy,Herdeiro:2019tmb,Brihaye:2019gla,Hod:2020ljo} (with the exception of \cite{MassTermMyung}, that studied the case of a configuration of the EMS model with a massive scalar field ($U(\phi) \sim \mu^2\phi^2$) only for a very specific BH charge to mass ratio). A more realistic treatment of the problem requires a full analysis of spontaneous scalarization in EMS models with a massive (and self-interacting) scalar field, which is the aim of this work. We extend previous works to inquire on the effects of a massive and self-interacting scalar-field ($U(\phi) \sim \mu^2 \phi^2 + \lambda \phi^4$) on generic EMS BHs for several coupling functions. Extended scalar-tensor theories arise naturally in string theory and broken supersymmetry leads to massive scalar fields. The inclusion of scalar field mass suppresses the scalar field at the length scale of the order of the Compton wavelength, which may help in reconciling the theory with observations for a much broader range of the coupling parameters and functions \cite{Doneva:2019vuh}. Such study of spontaneous scalarization with a massive and self-interacting scalar field was performed in \cite{Macedo:2019sem} in the context of eSTGB models, leading to two main conclusions: (i) a mass term for the scalar field alters the threshold for the onset of scalarization; (ii) the quartic self-coupling is sufficient to produce scalarized solutions that are stable against radial perturbations, without the need to resort to (exotic) higher-order terms in the Gauss-Bonnet coupling function. In this work we inquire if a parallelism between the results for eSTGB and EMS models emerge. Scalar field self-interactions have been further studied, \textit{e.g.}, in the context of Kerr BHs with synchronized hair \cite{Herdeiro:2015tia,Herdeiro:2016gxs}.

\par This paper is organized as follows. In Section \ref{S2} we present the model, obtain the field equations for the ansatz that describes the class of solutions of interest, providing some details on the construction of the solutions and, following Ref. \cite{MainPaper3}, propose a classification of the BH solutions, based on the behaviour of the coupling function. Then we describe how the emergence of the scalarized solutions is computed, in linear theory, comparing our results with the ones from  Ref.\cite{MassTermMyung} and discuss the effective potential for spherical perturbations and the S-deformation method. Section \ref{S3} contains the bulk of our results and the analysis of the two main examples considered (dilaton and scalarized cases), including the analysis of the domains of existence, studying the thermodynamic preference, the radial profiles of the solutions and the analysis of the solutions' radial stability. Final remarks are presented in Section \ref{S4}.


%
 \section{The model}\label{S2}
%
The action of the family of EMS models we wish to consider is given by \ref{eq:action}. We focus on the case where the scalar field is massive and self-interacts
\begin{equation}
U(\phi)/2=\mu^2 \phi^2 + \lambda \phi^4\ ,
\label{eq:potential}
\end{equation}
where $\mu$ is the scalar field mass and the self-coupling is positive, $\lambda>0$. The equations of motion that follow from \eqref{eq:action} are
\begin{equation}
R_{\mu \nu} - \frac{1}{2}g_{\mu \nu} R =2\left( T_{\mu \nu}^{(\phi)} + T_{\mu \nu}^{(EM)} \right),
\label{eq:EinsteinEq}
\end{equation}
\begin{equation}
\partial_{\mu} \left(\sqrt{-g}f(\phi)F^{\mu \nu} \right)=0,
\label{eq:MaxwellEq}
\end{equation}
\begin{equation}
\Box \phi = \frac{\dot f(\phi)F_{\mu \nu}F^{\mu \nu} + \dot U(\phi)}{4},
\label{eq:KleinGordonEq}
\end{equation}
where the dot denotes differentiation with respect to the scalar field, \textit{i.e.}, $\dot f(\phi) \equiv df/d\phi$ and the energy-momentum tensor is given by
\begin{equation}
T_{\mu \nu}^{(\phi)}=\partial_\mu \phi \partial_\nu \phi - \frac{1}{2}g_{\mu \nu}\left[ \partial_\alpha \phi \partial^\alpha \phi + \frac{1}{2}U(\phi) \right], \qquad T_{\mu \nu}^{(EM)} = f(\phi) \left( F_{\mu \alpha} F_{\nu}^{\alpha}-\frac{1}{4} g_{\mu \nu} F_{\alpha \beta}F^{\alpha \beta} \right).
\label{eq:EMtensor}
\end{equation}
A generic, static and spherically symmetric line element used to describe both scalar-free and scalarized solutions is
\begin{equation}
\label{ma}
    d s ^ { 2 } = - N ( r ) e ^ { - 2 \delta ( r ) } d t ^ { 2 } + \frac { d r ^ { 2 } } { N ( r ) } + r ^ { 2 } \left( d \theta ^ { 2 } + \sin ^ { 2 } \theta d \varphi ^ { 2 } \right) \ .
\end{equation}
Spherical symmetry requires the scalar field $\phi(r)$ to have a radial dependence only, and an electromagnetic 4-potential ansatz of the following type,
\begin{equation}
    A=V(r) dt.
    \label{aa}
\end{equation}
Functions $N,\delta,V,\phi$ have radial dependence only; for ease of notation this dependence will be omitted henceforth and a radial derivative will be denoted by a prime. With this ansatz, the equations of motion (\ref{eq:EinsteinEq},~\ref{eq:MaxwellEq},~\ref{eq:KleinGordonEq}) reduce to
\begin{equation}
\begin{aligned}
V'=\frac{Q}{r^2 f(\phi)}e^{-\delta} \ , \qquad \left(e^{-\delta} r^2 N \phi' \right)' = -\frac{1}{2} e^{\delta} r^2 \dot f(\phi) V'^2 + \frac{e^{-\delta}}{4} r^2 \dot U(\phi) \ ,\\
\delta'=-r\phi'^2 \ , \qquad N' = \frac{1-N(1-r\delta')}{r}-\frac{1}{2}rU(\phi) - \frac{Q^2}{r^3 f(\phi)} ,
\end{aligned}
\label{c4:eq:scalarfield}
\end{equation}
where $Q$ is an integration constant interpreted as the electric charge measured at infinity. To solve this set of coupled, non-linear ordinary differential equations, we have to implement suitable boundary conditions for the desired functions and corresponding derivatives. We assume the existence of an event horizon at $r=r_H>0$ and that the solution possesses a power series expansion in $(r-r_H)$
\begin{equation}
    \begin{array} { l } { N ( r ) = N _ { 1 } \left( r - r _ { H } \right) + \ldots, } \qquad { \delta ( r ) = \delta _ { 0 } + \delta _ { 1 } \left( r - r _ { H } \right) + \ldots, } \\ { \phi ( r ) = \phi _ { 0 } + \phi _ { 1 } \left( r - r _ { H } \right) + \ldots, } \qquad { V ( r ) = v _ { 1 } \left( r - r _ { H } \right) + \ldots \ . } \end{array}
    \label{nhe}
\end{equation}
Plugging these expansions in the field equations, the lower order coefficients are determined to be
\begin{equation}
    \begin{array} { l } { N_1=-\frac{Q^2 - r_H^2 f(\phi_0)}{r_H^3 f(\phi_0)} - \frac{1}{2} r_H U(\phi_0), } \qquad { \delta_1 = -\phi _1 ^{\ze \ze 2}\ze \ze r_H, } \\ { v_1=\frac{Q}{r_H^2 f(\phi_0)}e^{-\delta_0}, } \qquad { \phi_1 = \frac{2Q^2 \dot f(\phi_0) - r_H^4 f(\phi_0)^2 \dot U(\phi_0)}{4Q^2 r_H f(\phi_0) - 4r_H^3 f(\phi_0)^2 + 2r_H^5 f(\phi_0)^2 U(\phi_0)} \ . } \end{array}
\end{equation}
One observes that only two of the six parameters introduced in the expansions~\eqref{nhe} are independent, which we choose to be $\phi_0$ and $\delta_0$, the remaining being derived from these ones. The solutions in the vicinity of the horizon are determined by these two parameters, together with $(r_H,Q,\mu,\lambda)$. Some physical horizon quantities, such as  the Hawking temperature $T_H$, the horizon area $A_H$, the energy density $\rho(r_H)$ and the Kretschmann scalar $K(r_H)$, are then determined by these parameters as follows: 

\begin{align}
\begin{split}
 	 T_H=\frac{1}{4\pi} N_1 e^{-\delta_0}\ , \qquad
	A_H=4\pi r_H^2\ , \qquad 
	\rho(r_H) = \frac{2Q^2}{r_H^4 f(\phi_0)} + U(\phi_0)\ ,
\\
	 K(r_H)=\frac{4}{r_H^8 f(\phi_0)^2} \left( 5Q^4 - 6r_H^2 Q^2 f(\phi_0) +3 f(\phi_0)^2 r_H^4 \right) + 4U(\phi_0) \frac{\rho(r_H)r_H^2 - 2}{r_H^2} \ .\
\end{split}
\end{align}

\par To obtain the boundary conditions at spatial infinity one performs an asymptotic approximation of the solution in the far field. Then the equations of motion yield
\begin{equation}
    N ( r )=1 - \frac{2M}{r} + \frac { Q ^ { 2 } + Q _ { s } ^ { 2 } } { r^2 } + \ldots\ , \qquad \phi(r)=\frac{Q_{s}}{r} e^{-\mu r}+\ldots \, , \qquad V(r)=\Phi_e+\frac{Q}{r}+\ldots \, , 
\end{equation}
which introduce three new parameters: the scalar charge $Q_s$,  the electrostatic potential difference between the horizon and infinity $\Phi_e$ and the ADM mass $M$. From these asymptotic expansions one collects a set of nine parameters: $(r_H,Q,\mu,\lambda,\phi_0,\delta_0,Q_s,\Phi_e,M)$. As we shall see below, in section \ref{S3.1}, when analyzing the domains of existence of solutions, the full integration of the field equations relates these parameters, and, for each choice of the coupling functions, the solutions of interest actually form a family of solutions with only 4 parameters, typically taken to be the global charges $(M,Q)$ and the self-coupling parameters $(\mu,\lambda)$.
For later use we gather the following results and definitions:
\begin{gather}
    q\equiv \frac{Q}{M}\ ,\, \qquad a_H\equiv \frac{A_H}{16\pi M^2} \ , \qquad \hat \mu \equiv Q \mu \, , \qquad \hat \lambda \equiv Q^2 \lambda \ , \qquad \hat{Q}_s \equiv \frac{Q_s}{Q},
\end{gather}
where $\hat \mu$ and $\hat \lambda$ are dimensionless self-coupling parameters, $q$ is the \textit{reduced} charge, $a_H$ is the \textit{reduced} horizon area and $\hat Q_s$ the \textit{reduced} scalar charge. These reduced quantities are convenient because they are invariant under the scaling symmetry
\begin{equation}
    r \to \eta r\ , \qquad \xi \to \eta \xi \ , \qquad \mu \to \mu/\eta \ , \qquad \lambda \to \lambda/\eta^2 \ ,
\end{equation}
where $\eta$ is a constant and $\xi$ represents any of the global charges of the model.

\subsection{Physical relations and tests to the code}
\par To solve the equations of motion \eqref{c4:eq:scalarfield}, which take the form of four coupled ordinary differential equations we apply a Runge-Kutta strategy given the aforementioned boundary conditions. Our numerical method implements a six(five) Runge-Kutta integration algorithm (RK65) with an adaptative step size and a shooting method. The latter is implemented in the unknown parameters $\phi_0$ and $\delta_0$, and ensures the fulfillment of the boundary conditions. This code is written in C and was developed and extensively tested by us, being previously used with success in the works \cite{MainPaper1,MainPaper2,MainPaper3,MainPaper4}.
\par Let us now briefly consider two physical relations that, besides their physical content, are used to test the accuracy of the solutions found numerically. These are a Smarr-type law and a virial-type relation.
%
%
\subsubsection{A Smarr-type law}
\par The Smarr law \cite{4lawsBH,Smarr} provides a relation between the total mass of the spacetime and other measurable quantities, like the horizon temperature and area. Its information complements that of the equations of motion, making it an interesting test to assess the accuracy of BH solutions obtained numerically. 

The Smarr law can be obtained via the integral mass formula, that for our model reads
\begin{equation}
    M=\frac{1}{2}T_H A_H - \frac{1}{16\pi} \int_V (2T^\nu_\mu - T \delta^\nu_\mu) k^\mu d\Sigma_\nu \ ,
\end{equation}
where $k^\mu$ is the Killing vector field associated to staticity and $T$ is the trace of the energy-momentum tensor defined in \eqref{eq:EMtensor}. One can thus arrive at the Smarr-type law
\begin{equation}
    M=\frac{1}{2}T_H A_H + \Phi_e Q - \frac{1}{2}\int_{r_H}^\infty r^2 e^{-\delta} U(\phi) dr \ .
\end{equation}

\subsubsection{A virial-type relation}
\par Scaling arguments, initiated by the work of Derrick~\cite{Derrick:1964ww}, are a powerful tool to establish no-go theorems for solitonic solutions (see $e.g.$~\cite{Herdeiro:2019oqp}),  no-hair theorems for BH solutions \cite{ScalarHairReview}, as well as to provide a physical relation that must be obeyed by solutions of a given model. These relations are generalisations of the canonical virial theorem, that states an energy balance, and are often described as virial relations. They are typically independent from the equations of motion; thus, again, they are useful in assessing the accuracy of numerically generated solutions.

Consider the effective action
\begin{equation}
    S_{\rm eff} = \int_{r_H}^{\infty} dr \mathcal{L}_{\rm eff}\ ,
\end{equation}
where
\begin{equation}
	\mathcal{L}_{\rm eff}=\frac{e^{\delta}}{2} r^2 f(\phi) V'^2 + \frac{e^{-\delta} }{2} \left(1-N-rN'-r^2 N \phi'^2 - \frac{1}{2}r^2 U(\phi)\right) \ ,
\end{equation}
is the effective Lagrangian that can be obtained from the action \eqref{eq:action} by integrating the trivial angular dependence. Now assume that a charged BH solution with scalar hair exists, described by the functions $\phi(r),\delta(r),V(r),N(r)$, with suitable boundary conditions at the event horizon and at infinity. Next, consider the 1-parameter family of configurations described by the scaled functions 
\begin{equation}
    F_\lambda(r) \equiv F(r_H+\lambda(r-r_H))\ ,
\end{equation}
with $F \in \{\phi,\delta,V,N\}$. If the initial configuration was indeed a solution, then the effective action for the scaled configurations must possess a critical point at $\lambda=1$:  $\left({dS_{eff}^\lambda}/{d\lambda}\right)_{\lambda=1}=0$. From this condition one obtains the virial-type relation 
\begin{equation}
    \int_{r_{H}}^{\infty} d r\left\{ e^{-\delta} r^{2} {\phi^\prime}^2\left[1-\frac{r_{H}}{r}\left(1+N\right)\right]\right\} = \Phi_e Q + \int_{r_H}^{\infty} d r \left\{\frac{2r_H}{r}Q V^\prime + e^{-\delta}r^2 U(\phi) \left(\frac{r_H}{r} -\frac{3}{2}\right)\right\}\ .
\end{equation}
\par One can show that the left hand side integrand is strictly positive and that the integral over the scalar field potential on the right hand side is strictly negative (assuming $\mu^2\geq 0$ and $\lambda \geq 0$). Thus, the virial identity shows that a nontrivial scalar field requires a nonzero electric charge so that the right hand side is nonzero. As an immediate corollary, neutral BHs cannot be hairy in this model. 

\par We remark that throughout this work, solutions are well within the numerical errors: our tests have exhibited a relative difference of the order of $10^{-7}$ for the virial relation and $10^{-6}$ for the Smarr relation.

%
\subsection{Classification of EMS models}
%
\par Depending on the choice of coupling $f(\phi)$, the RN BH may or may not be a solution of the equations of motion of the model, which can be better seen from the scalar field equation of motion \eqref{eq:KleinGordonEq}. This leads to two classes of EMS models \cite{MainPaper3}.
\subsubsection{Class I - Models without a scalar-free solution}
\par In this class of EMS models $\phi(r)=0$ does not solve the field equations and so, the RN BH is not a solution. From the scalar field equation of motion \eqref{eq:KleinGordonEq}
\begin{equation}
\dot f(0) \neq 0 \ .
\end{equation}
Such representative coupling is the standard dilatonic coupling (studied in the massless, non-self-interacting case in \cite{MainPaper3,DilatonGibbons,DilatonGibbons2,DilatonGarfinkle}).
\begin{equation}
f(\phi)=e^{\alpha \phi}
\end{equation}
in which case we refer to $\phi$ as a dilaton field. Three reference values for the dilaton coupling constant $\alpha$ are \cite{MainPaper3}: $\alpha=0$ (Einstein-Maxwell theory), $\alpha=2$ (low energy strings), $\alpha=2\sqrt{3}$ (Kaluza-Klein theory). Massive dilaton studies have been conducted in \cite{Gregory:1992kr,Horne:1992bi}.

\subsubsection{Class II - Models with a scalar-free solution}
\par In this class of EMS models $\phi(r)=0$ solves the field equations and so, the RN BH is a solution \footnote{For simplicity we assume $f(0) = 1$ to identify the scalar-free solution with the standard RN BH.}. From the scalar field equation of motion \eqref{eq:KleinGordonEq}
\begin{equation}
\dot f(0)=0 \ .
\end{equation}
The RN solution, however, is (in general) not unique. These EMS models may contain a new set of BH solutions, with a non-trivial scalar field profile - the scalarized BHs. As discussed in \cite{MainPaper3}, such new set of BH solutions may exist in models where the RN BH is (class IIA) or is not (class IIB) unstable. In this work we are interested in the first (IIA), for which \textit{spontaneous scalarization} occurs. The second case is studied in great detail in \cite{HotColdBald}. The spontaneously scalarized (hereby dubbed ``\textit{scalarized}") BHs bifurcate from RN BHs, and reduce to the latter for $\phi=0$. This bifurcation moreover, may be associated to a tachyonic instability, against scalar perturbations $\delta \phi$, of the RN BH. These obey
\begin{equation}
(\Box - \mu_{eff}^2) \delta \phi = 0,
\label{eq:scalarfieldlinear}
\end{equation}
with $\mu_{eff}^2<0$ given by
\begin{equation}
	\mu_{eff}^2=-\frac{\ddot f(0) Q^2}{2r^4} + \mu^2,
	\label{eq:mueff}
\end{equation}
being unaffected by the self-coupling $\lambda$. Such representative coupling that will be studied in greater detail later is
\begin{equation}
f(\phi)=e^{\alpha \phi^2} \ .
\end{equation}
The coupling constant $\alpha$ is taken as a positive. This coupling was studied in the massless, non-self-interacting case in detail in \cite{MainPaper1,MainPaper2,MainPaper3}.\\
\paragraph{Bifurcation of solutions: the existence line}\mbox{}\\
\par Let us now consider the onset of spontaneous scalarization. We assume that the model under consideration admits the RN BH of Einstein-Maxwell theory as the scalar-free solution, that is~\eqref{ma}-\eqref{aa} with
\begin{equation}
\delta=0 \ , \qquad N(r)=1-\frac{2M}{r}+\frac{Q^2}{r^2} \ , \qquad V(r)=\frac{Q}{r}\ .
\label{eq:RNsolution}
\end{equation}
The scalarization phenomenon is assessed by considering scalar perturbations of the RN solution within the considered model.  Following~\cite{MainPaper1,MainPaper2,MainPaper3,MainPaper4}, we take a spherical harmonics decomposition of the scalar field perturbation:
\begin{equation}
    \delta \phi = \sum_{\ell,\mathbf{m}} Y_{\ell,\mathbf{m}}(\theta,\varphi) U_\ell(r) \ .
\end{equation}
With this ansatz, the scalar field equation of motion \eqref{eq:scalarfieldlinear} simplifies to
\begin{equation}
\label{sfep}
    \frac { e ^ { \delta } } { r ^ { 2 } } \left( \frac { r ^ { 2 } N } { e ^ { \delta } } U _ { \ell } ^ { \prime } \right) ^ { \prime } - \left[ \frac { \ell ( \ell + 1 ) } { r ^ { 2 } } + \mu _ { \mathrm { eff } } ^ { 2 } \right] U _ { \ell } = 0 \ ,
\end{equation}
considering the background solution as the RN BH of Einstein-Maxwell theory \eqref{eq:RNsolution}, we obtain for the perturbation equation \eqref{sfep}
\begin{equation}
\left[ \left(r ^ { 2 }-2Mr+Q^2\right)  U _ { \ell } ^ { \prime } \right] ^ { \prime } - \left[\ell ( \ell + 1 ) + r^2 \mu _ { \mathrm { eff } } ^ { 2 } \right] U _ { \ell } = 0 \ .
\end{equation}
Recall that in order for a tachyonic instability to settle in, we must have $\mu_{eff}^2 <0$. Once the coupling functions are fully fixed, solving~\eqref{sfep} is an eigenvalue problem: for a given $\alpha$ and $\ell$, requiring an asymptotically vanishing, regular at the horizon, smooth scalar field, a discrete set of BHs solutions are selected, $i.e.$ a discrete set of RN solutions, each with a certain reduced charge $q$. These are the \textit{bifurcation points} from the scalar-free solution. They are labelled by an integer $n\in \mathbb{N}_0$; $n=0$ is the fundamental mode, whereas $n\geqslant 1$ are excited states (overtones). The RN solutions with a smaller (larger) $q$ than that of the bifurcation point are stable (unstable) against the corresponding scalar perturbation. In particular, the first bifurcation point, $i.e.$, the one with the smallest $q$, which corresponds to the mode $\ell=0$ and $n=0$, marks the onset of the scalarization instability. Only RN BHs with $q$ smaller than the first bifurcation point are stable against any sort of scalar perturbation. Then, a scalarized solution can be dynamically induced by a scalar perturbation of the background, as long as the scalar-free RN solution is in the unstable regime.

\begin{figure}[ht!]
\centering
\includegraphics[width=.5\textwidth]{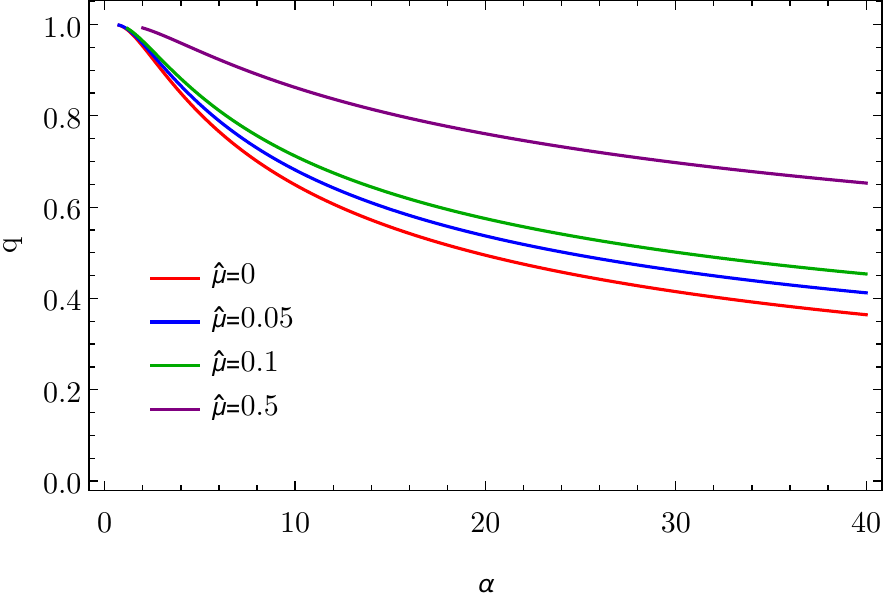}
\caption{Set of bifurcating points - the existence lines - of scalarized solutions in the $(\alpha,q)$ plane for a sample of $\hat \mu$ values. Higher $\hat \mu$ values lead to higher required $q$ for bifurcation, altering the threshold for the onset of scalarization.}
\label{fig:s-existencelines}
\end{figure}

At each bifurcation point, a new family of (fully non-linear) scalarized BH solutions emerges from the RN family, as static solutions of the equations of motion of the full model. In this paper we shall consider only the first bifurcation point and the corresponding new family of spherically symmetric scalarized BHs that bifurcate from the RN family as overtone solutions are expected to be unstable \cite{EMS-Stability,MassTermMyung}. The existence line (the set of bifurcating points from the scalar-free solution) is altered by the scalar-field mass because it has a suppressing effect for the tachyonic instability, altering the threshold for the onset of scalarization as seen in Fig. \ref{fig:s-existencelines} where the existence lines are plotted for a sample of $\hat \mu$ values. Besides shifting the minimum value of $\alpha$ to higher values, for each constant $\alpha$ line, a higher value of $\hat \mu$ implies a higher charge to mass ratio $q$ for bifurcation.

Our values for the bifurcation points match with great precision the ones obtained in Table I of \cite{MassTermMyung}, for a study of scalarized BHs of the same model considered in this work, with a specific value of $q=0.7$ and a scalar-field mass term $\mu^2=\alpha/\beta$, where $\beta$ is a mass-like parameter. In ref. \cite{MassTermMyung} the results for the minimum values of $\alpha$ for bifurcation for a RN BH with $q=0.7$ are presented in such a way that they depend on the parameter $\beta$, which is not scale invariant. Table \ref{tab:bifq07} presents the same results but with the minimum values of $\alpha$ for bifurcation depending on $\hat \mu$, a scale invariant quantity, so that these results can be applied to any system in consideration.

\begin{table}[ht!]
    \centering
    \begin{tabular}{c|c|c|c|c|c|c|c}
        $\hat \mu$ & 0.0 & 0.0228 & 0.0365 & 0.05 & 0.10 & 0.365 & 0.50\\ \hline
        $\alpha$ & 8.019 & 8.493 & 8.82 & 9.18 & 10.69 & 21.80 & 29.52
    \end{tabular}
    \caption{Bifurcation points ($n=0$ mode) for several values of $\hat \mu$ for a RN BH with $q=0.7$. The results match the ones from \cite{MassTermMyung}.} 
    \label{tab:bifq07}
\end{table}

%
\subsection{Effective potential for spherical perturbations and stability}\label{subsec:effpot}
%
\par Let us also introduce a diagnosis analysis of perturbative stability, against spherical perturbations, that shall be applied to the solutions derived and discussed in the next section. Following a standard technique, see $e.g.$ \cite{MainPaper4}, we consider spherically symmetric, linear perturbations of an equilibrium solution, keeping the metric ansatz, but allowing the functions $N,\delta,\phi,V$ to depend on $t$ as well as on $r$:
\begin{equation}
    ds^2=- \tilde N(r,t)e^{-2 \tilde \delta(r,t)} dt^2+\frac{dr^2}{\tilde N(r,t)}+r^2(d\theta^2+\sin^2 \theta d\varphi^2) \ ,
 \qquad A= \tilde V(r,t) dt \ , \qquad \phi=\tilde \phi(r,t) \ .
\end{equation}
The time dependence enters as a Fourier mode with frequency $\Omega$, for each of these functions:
\begin{eqnarray}
&&
 \tilde N(r,t)=N(r)+\epsilon N_1(r)e^{-i \Omega t}\ , \qquad  \tilde \delta(r,t)=\delta(r)+\epsilon \delta_1(r)e^{-i \Omega t} \ , 
\\
&&
\nonumber
 \tilde \phi(r,t)=\phi(r)+\epsilon \phi_1(r)e^{-i \Omega t}\ , \qquad \tilde V(r,t)=V(r)+\epsilon V_1(r)e^{-i \Omega t}\ .
\end{eqnarray}
From the linearized field equations around the background solution, the metric perturbations and $V_1(r)$ can be expressed in terms of the scalar field perturbation,
\begin{eqnarray}
N_1=-2r N\phi' \phi_1 \ , \qquad \delta_1'=-2 r\phi' \phi_1' \ , \qquad V_1'= -V'\left( \delta_1 + \frac{\dot f(\phi)}{f(\phi)}\phi_1 \right) \ ,
\end{eqnarray}
thus yielding a single perturbation equation for $\phi_1$. Introducing a new variable $\Psi(r)=r\phi_1$, the scalar-field equation of motion may be written as
\begin{equation}
    \left(N e^{-\delta}\right)^2 \Psi'' + N e^{-\delta}\left(N e^{-\delta}\right)' \Psi' + \left(\Omega^2 - U_\Omega\right) \Psi = 0 \ ,
\end{equation}
which, by introducing the 'tortoise' coordinate $x$ as $dx/dr=e^{\delta}/N$ \cite{qnmcardoso}, can be written in the standard one-dimensional Schr\"odinger-like form:
\begin{equation}
    -\frac{d^2 }{dx^2}\Psi+U_{\Omega} \Psi=\Omega^2 \Psi \ .
    \label{c4:eq:stab-shrod}
\end{equation}
The effective potential that describes spherical perturbations $U_{\Omega}$ is defined as:
\begin{equation}
    U_\Omega = U_{SI} + U_0 \ ,
\end{equation}
with the subscripts $SI$ and $0$ standing for ``Self-Interactions" and ``No-Self-Interactions" respectively, and
\begin{align}
    \begin{split}
    	U_{SI}&=\frac{e^{-2\delta}N}{2} \Big\{ \mu^2 \left[1+4r\phi \phi' + \phi^2 \left(-1+2r^2\phi'^2 \right) \right] + \lambda \phi^2 \left[6+8r\phi \phi' + \phi^2 \left(-1+2r^2 \phi'^2 \right)  \right] \Big\} \ , \\
        U_0&=\frac{e^{-2\delta}N}{r^2} \left\{ 1 - N -2r^2\phi'^2 - \frac{Q^2}{r^2 f(\phi)} \left( 1-2r^2\phi'^2 + \frac{\ddot f(\phi)}{2 f(\phi)} + 2r\phi' \frac{\dot f(\phi)}{f(\phi)}-\left(\frac{\dot f(\phi)}{f(\phi)}\right)^2 \right) \right\} \ .
    \end{split}
    \label{eq:effectivepot}
\end{align}
An unstable mode would have $\Omega^2<0$, which for the asympotic boundary conditions of our model is a bound state. It follows from a standard result in quantum mechanics (see $e.g.$~\cite{Messiah:1961}), however, that eq. (\ref{c4:eq:stab-shrod}) has no bound states if $U_\Omega$ is everywhere larger than the lowest of its two asymptotic values, $i.e.$, if it is positive in our case. 
Thus an everywhere positive effective potential proofs mode stability against spherical perturbations.

We remark that the existence of a region of negative potential is a necessary but not sufficient condition for instabilities to be present. In fact, for the fundamental, spherically symmetric scalarized solutions in~\cite{MainPaper1,MainPaper2} with $\hat \mu=0$ and $\hat \lambda=0$, this region occurs for some solutions near the existence line, which are, nonetheless, stable~\cite{EMS-Stability}. As will later be observed, some black hole solutions of the model in study contain a potential well in the effective potential. To assess the stability of such solutions we resort to the S-deformation method \cite{Kimura:2017uor,Kimura:2018eiv,Kimura:2018whv,HotColdBald}. We shall briefly review the procedure and refer the reader to \cite{Kimura:2017uor,HotColdBald} for a more detailed explanation. Multiplying \eqref{c4:eq:stab-shrod} by $\bar{\Psi}$ (where the bar denotes complex conjugation) and integrating from the horizon to infinity we obtain
\begin{equation}
-\left[\bar{\Psi} \frac{d \Psi}{d x}\right]_{-\infty}^{\infty}+\int_{-\infty}^{\infty} d x\left[\left|\frac{d \Psi}{d x}\right|^{2}+U_\Omega|\Psi|^{2}\right]=\Omega^{2} \int_{-\infty}^{\infty} d x|\Psi|^{2},
\label{eq:shrod2}
\end{equation}
where the first term on the left hand side vanishes due to the imposed boundary conditions  on $\Psi$ \footnote{Here we impose that $\Psi$ and $d\Psi/dx$ vanish as $x \to \pm \infty$ and are continuous and bounded everywhere \cite{Kimura:2017uor}.}. It follows that, as previously stated, $ U_{\Omega} >0$ implies $\Omega^2 >0$. Now we generalize \eqref{eq:shrod2} by introducing an arbitrary function $S$ (deformation function) into the wave equation obtaining \footnote{Where we made use of the identity $$-\bar{\Psi} \frac{d^{2} \Psi}{d x^2}=-\frac{d}{d x}\left(\bar{\Psi} \frac{d \Psi}{d x}\right)+\left|\frac{d \Psi}{d x}\right|^{2}$$}
\begin{equation}
-\bar{\Psi} \frac{d^{2} \Psi}{d x^2}+U_{\Omega}|\Psi|^{2} \equiv -\frac{d}{d x}\left(\bar{\Psi} \frac{d \Psi}{d x}+S|\Psi|^{2}\right)+\left|\frac{d \Psi}{d x}+S \Psi\right|^{2} +\left(U_{\Omega}-S^{2}+\frac{d S}{d x}\right)|\Psi|^{2}=\Omega^{2}|\Psi|^{2}.
\label{eq:shrod3}
\end{equation}
Integrating \eqref{eq:shrod3} and imposing that $S$ is smooth everywhere and does not diverge at the boundaries, together with the boundary conditions for $\Psi$ yields
\begin{equation}
\int_{-\infty}^{\infty} d x\left|\frac{d \Psi}{d x}+S \Psi\right|^{2}+\int_{-\infty}^{\infty} d x\left(U_{\Omega}-S^{2}+\frac{d S}{d x}\right)|\Psi|^{2}=\Omega^{2} \int_{-\infty}^{\infty} d x|\Psi|^{2}.
\end{equation}
The first term on the left hand side is strictly positive, thus if there exists a function $S$ such that the second term on the left hand side vanishes or is positive, we conclude that no modes with $\Omega^2<0$ exist. Establishing stability then reduces to showing that there exists a well behaved function $S$ such that
\begin{equation}
\frac{d S}{d x}=S^{2}-U_{\Omega} \Leftrightarrow \frac{dS}{dr} = \frac{S^2-U_\Omega}{N e^{-\delta}}.
\label{eq:sdef}
\end{equation}
In the next section we solve Eq. \eqref{eq:sdef} numerically for the cases where the effective potential yield negative regions, imposing the boundary condition $S(r_H)=0$. These potential wells occur for solutions near the existence line as was already observed in the $\hat \mu = \hat \lambda = 0$ case \cite{MainPaper2}. If the solutions are unstable, it is expected that when solving Eq. \eqref{eq:sdef} the $S$ function develops singularities as it happens, \textit{e.g.}, for solutions of the first branch of Ref. \cite{HotColdBald}.

%
\section{Numerical results for the Black Hole solutions}\label{S3}
%
\par In this section we present, for both the dilatonic and scalarized couplings, the main results, namely the effects of the mass term and self-interactions on the domains of existence, radial function profiles, thermodynamic preference and effective potential for spherical perturbations of the BH solutions.
\subsection{Domain of Existence}
\label{S3.1}
\par
\begin{figure}[ht!]
\centering
\includegraphics[width=.5\textwidth]{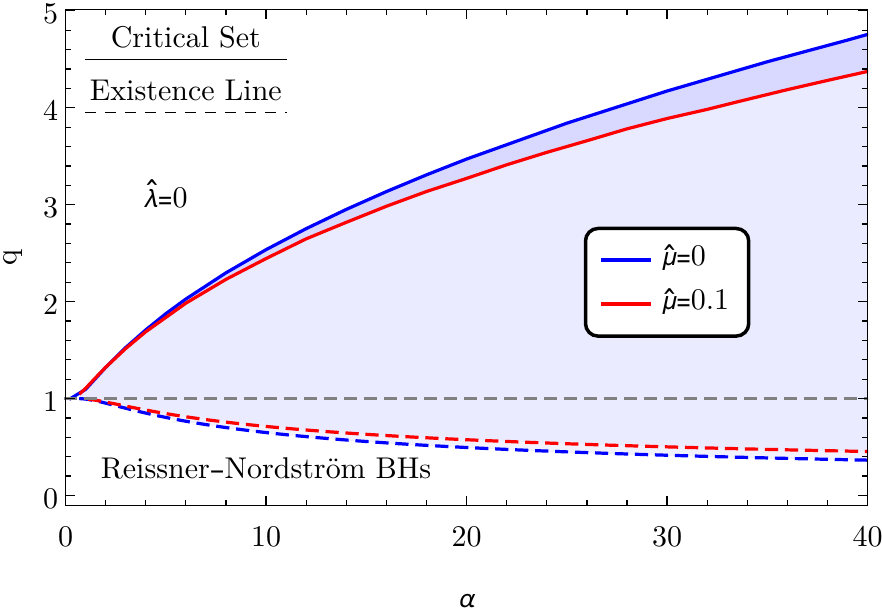}\hfill
\includegraphics[width=.5\textwidth]{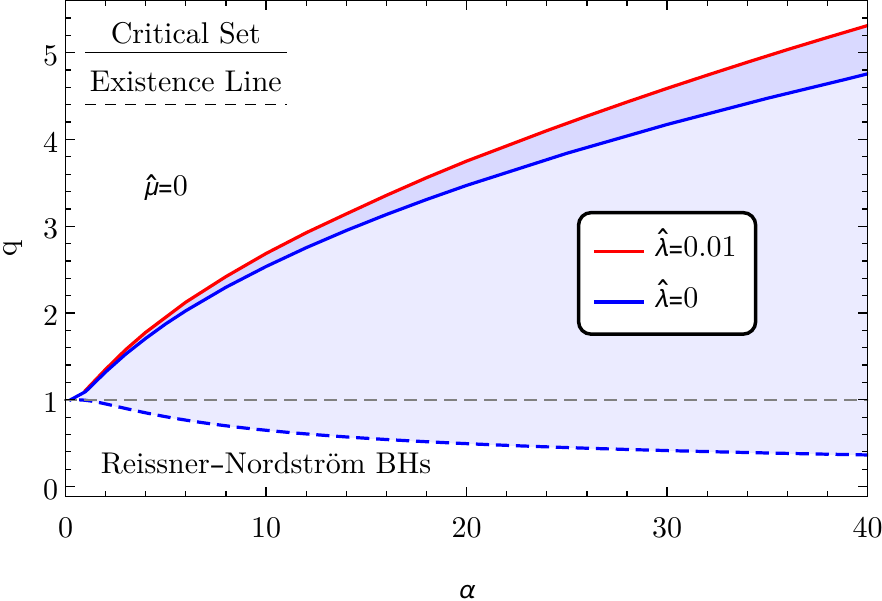}
\caption{Domain of existence of scalarized solutions in the $(\alpha,q)$ plane. Left: $\hat \lambda=0$ with $\hat \mu=0$ and $\hat \mu=0.1$ cases. The mass term has a narrowing effect on the domain of existence. Right: $\hat \mu=0$ with $\hat \lambda=0$ and $\hat \lambda=0.01$ cases. The self-interaction effects widen the domain of existence.}
\label{fig:s-domain}
\end{figure}

\begin{figure}[ht!]
\centering
\includegraphics[width=.5\textwidth]{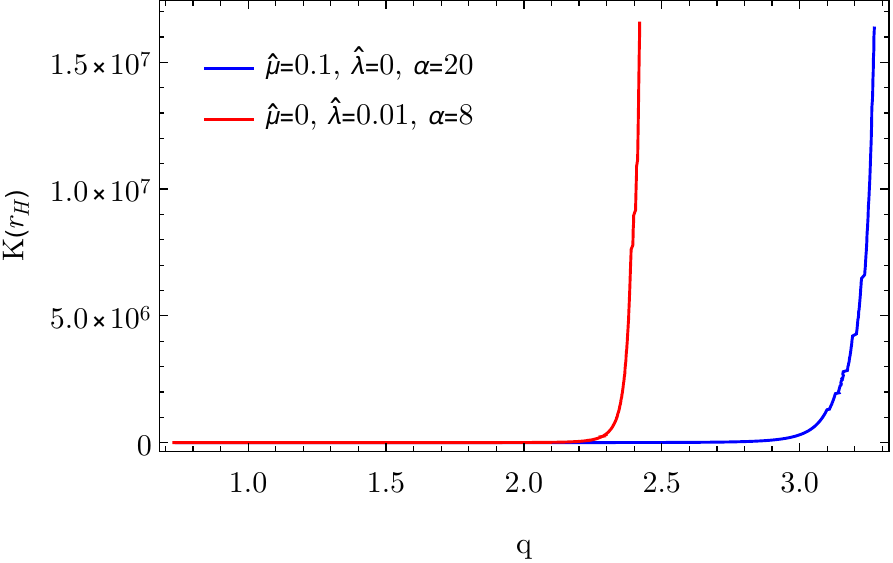}
\caption{Kretschmann scalar at the horizon $K(r_H)$ for the scalarized black hole configurations as a function of the charge to mass ratio $q$ for two different scenarios. $K(r_H)$ diverges as we approach the critical set.}
\label{fig:kretsch}
\end{figure}

 For the scalarized case, the domain of existence was obtained in the massless, non-self-interacting case in \cite{MainPaper1,MainPaper2,MainPaper3}, being bounded by an existence line and (in the absence of a magnetic charge) by a critical line, at which BH solutions are singular - numerics suggest a divergence of the Kretschmann scalar at the horizon and that $A_H \to 0$. The existence line will change depending on the scalar-field mass and is independent of the self-coupling $\lambda$ as previously observed in Fig. \ref{fig:s-existencelines}. The domain of existence for the scalarized coupling is presented in Fig. \ref{fig:s-domain}.
As seen in Fig. \ref{fig:s-domain} (left panel), the scalar field mass term has a narrowing effect on the domain of existence of scalarized solutions, because higher reduced charge $q$ is required for bifurcation and because overcharging is restricted - the critical set occurs for smaller $q$ values as compared to the massless case. On the other hand, the self-interaction effects widen the domain of existence - \textit{c.f.} Fig. \ref{fig:s-domain} (right panel). The existence line remains unchanged, while the critical set occurs for higher charge to mass ratios, suggesting a stabilizing effect. The divergence of the Kretschmann scalar at the critical set can be observed in Fig. \ref{fig:kretsch} for two different scenarios.

\par
\begin{figure}[ht!]
\centering
\includegraphics[width=.5\textwidth]{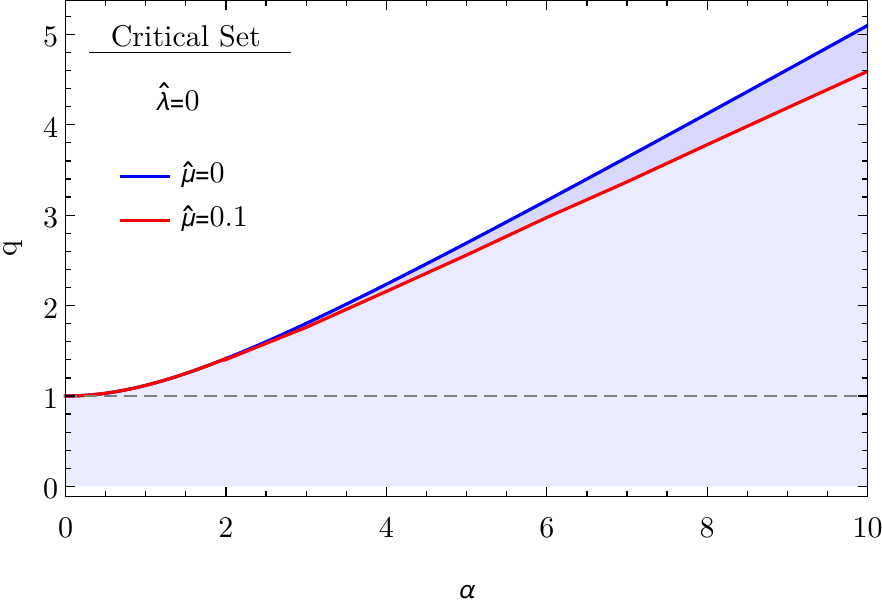}\hfill
\includegraphics[width=.5\textwidth]{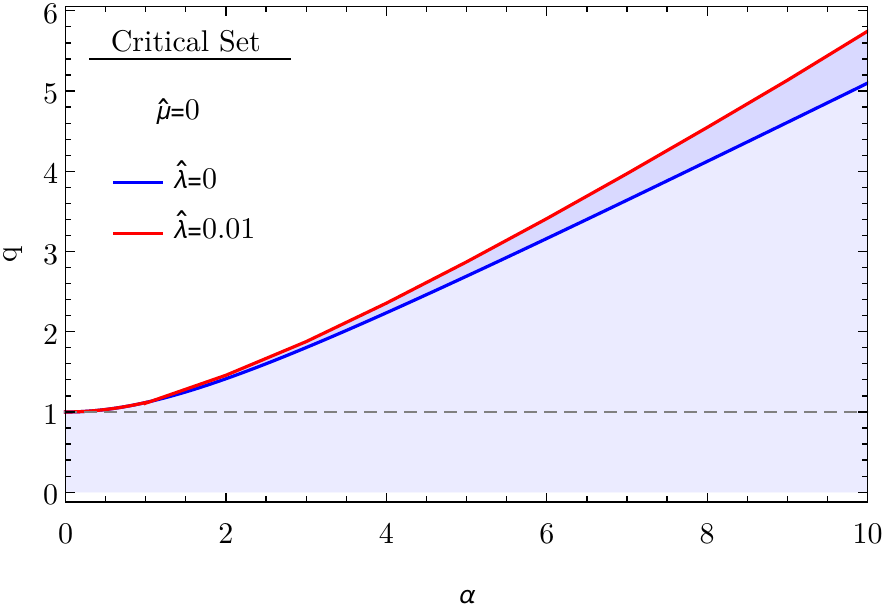}
\caption{Domain of existence of dilaton solutions in the $(\alpha,q)$ plane. Left: $\hat \lambda=0$ with $\hat \mu=0$ and $\hat \mu=0.1$ cases. The mass term has a narrowing effect on the domain of existence. Right: $\hat \mu=0$ with $\hat \lambda=0$ and $\hat \lambda=0.01$ cases. The self-interaction effects widen the domain of existence.}
\label{fig:d-domain}
\end{figure}

 Concerning the dilaton case, the domain of existence was obtained in the massless, non-self-interacting case in \cite{MainPaper3}, being bounded (in the absence of a magnetic charge) by a critical line. Note that there is no existence line since the model does not allow a scalar-free solution. Similarly to the scalarized case, as seen in Fig. \ref{fig:d-domain} (left panel), the scalar field mass term has a narrowing effect on the domain of existence of dilatonic solutions - the critical set occurs for smaller $q$ values as compared to the massless case. On the other hand, the self-interaction effects widen the domain of existence - \textit{c.f.} Fig. \ref{fig:d-domain} (right panel).

\par In either case, for each choice of fundamental parameters $(\mu,\lambda)$ and coupling constant $\alpha$, the domain of existence of scalarized solutions is fully determined by the global charges $(Q,M)$. In other words, a given family of solutions is totally described by the fundamental parameters (and global charges) $(\mu, \lambda, Q, M)$ while the remaining parameters of the model $(r_H,\phi_0,\delta_0,Q_s,\Phi_e)$ are fully determined by the fundamental ones (plus the global charges). For models with non-vanishing $\hat \mu$ and $\hat \lambda$, we can characterize their competition as a sum of individual effects as observed in Table \ref{tab:domain}.

\begin{table}[ht!]
    \centering
    \begin{tabular}{|c||c|c|c|c|c|c|c|c}\hline
        $\hat \mu$ & 0.0 & 0.0 & 0.1 & 0.1 & 0.1 & 0.1 & 0.05\\ \hline
        $\hat \lambda$ & 0.0 & 0.01 & 0.0 & 0.01 & 0.05 & 0.1 & 0.1\\ \hline \hline
        $q_{max}$ & 2.53 & 2.68 & 2.44 & 2.48 & 2.62 & 2.74 & 3.05\\ \hline
    \end{tabular}
    \caption{Maximum values allowed for the charge to mass ratio $q_{max}$ of scalarized black holes with $\alpha=10$ for a sample of self-interaction parameters.} 
    \label{tab:domain}
\end{table}

\subsection{Thermodynamic preference}

\par
\begin{figure}[ht!]
\centering
\includegraphics[width=.5\textwidth]{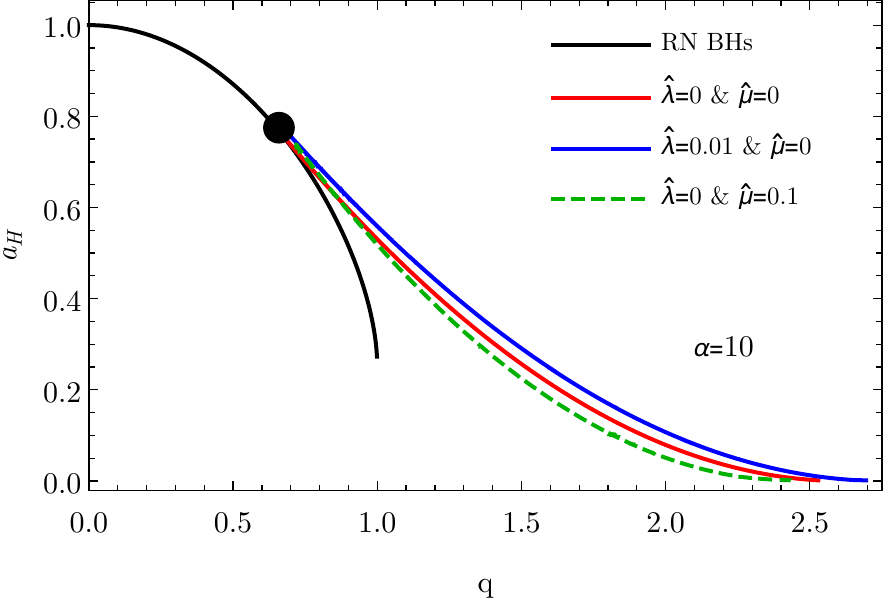}\hfill
\includegraphics[width=.5\textwidth]{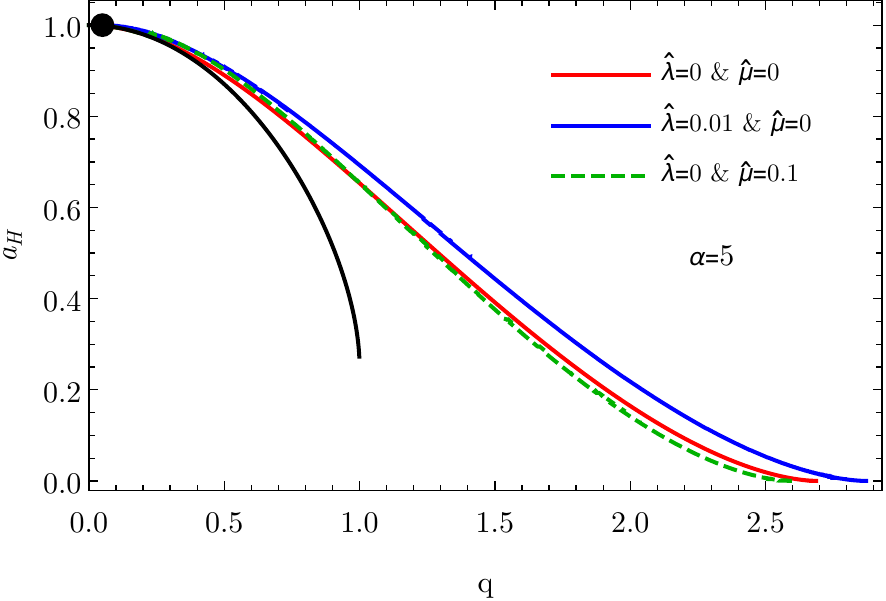}
\caption{$a_H$ $vs.$ $q$. (Left) The black line represents scalar-free RN BHs, while the red, blue and green lines are sequences of (numerical data points representing) scalarized BHs for $\alpha=10$. (Right) The red, blue and green lines represent sequences of (numerical data points representing) dilaton BHs for $\alpha=5$.}
\label{fig:entropy}
\end{figure}

 Concerning thermodynamic preference, since the model under consideration is General Relativity minimally coupled to some matter, the Bekenstein-Hawking BH entropy formula holds. Thus, the entropy analysis reduces to the analysis of the horizon area. It is convenient to use the reduced event horizon area $a_H$. In the region where the RN BHs and scalarized BHs co-exist - the non-uniqueness region -, for the same $q$, the scalarized solutions are always entropically preferred as seen in Fig. \ref{fig:entropy} (left). Entropic considerations, however, are not sufficient to establish if the endpoint of the instability of the RN BH is the corresponding hairy BH with the same $q$. Fully non-linear dynamical evolutions are required and, in the massless and non-self-interacting case, show that this is indeed the case for small enough charge to mass ratio or coupling $\alpha$, while for larger values of $q$, the endpoint would be a scalarized BH with smaller charge to mass ratio value than the unstable RN BH \cite{MainPaper1,MainPaper2}. We expect the same to occur for a massive and self-interacting scalar field. Such simulations are beyond of the scope of this work but represent a possible avenue of further research.

\subsection{Scalar field radial profiles}
\par
\begin{figure}[ht!]
\centering
\includegraphics[width=.5\textwidth]{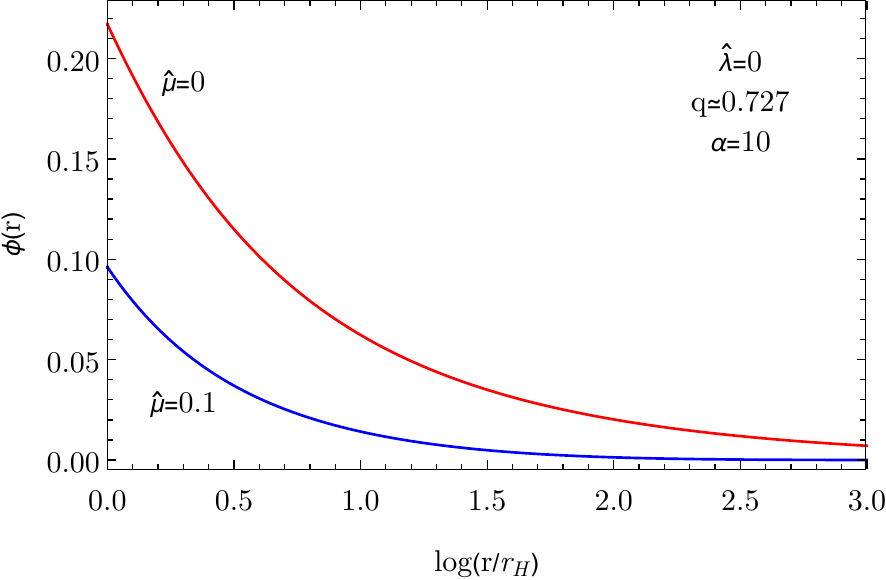}\hfill
\includegraphics[width=.5\textwidth]{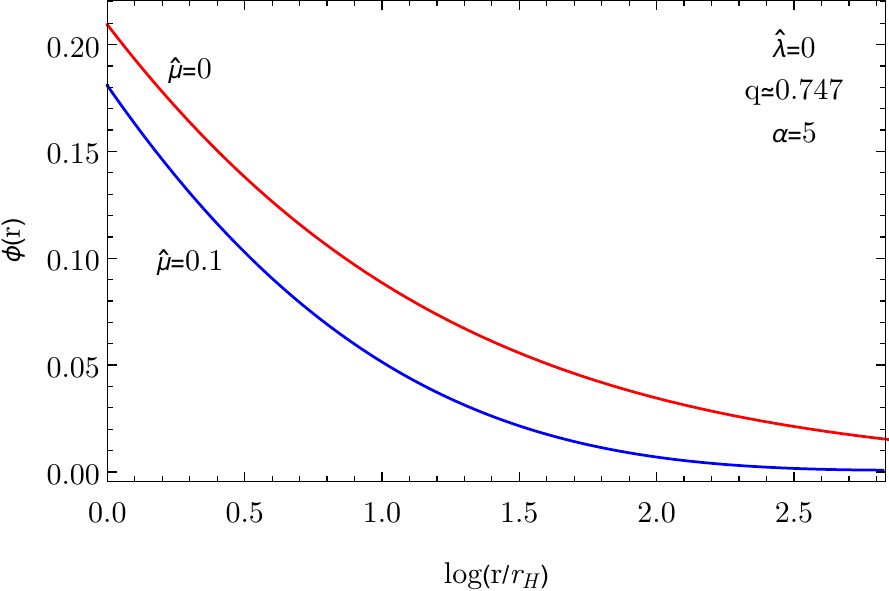}
\caption{(Left) Scalar field radial profiles for the scalarized case with $\hat \mu=0$ and $\hat \mu = 0.1$ while $q\approx 0.727$, $\alpha =10$ and $\hat \lambda=0$. (Right) Scalar field radial profiles for the dilatonic case with $\hat \mu=0$ and $\hat \mu = 0.1$ while $q\approx 0.747$, $\alpha=5$ and $\hat \lambda=0$. The mass term leads to a scalar field radial profile more concentrated in the neighborhood of the event horizon.}
\label{fig:radprofiles}
\end{figure}

\begin{figure}[ht!]
\centering
\includegraphics[width=.5\textwidth]{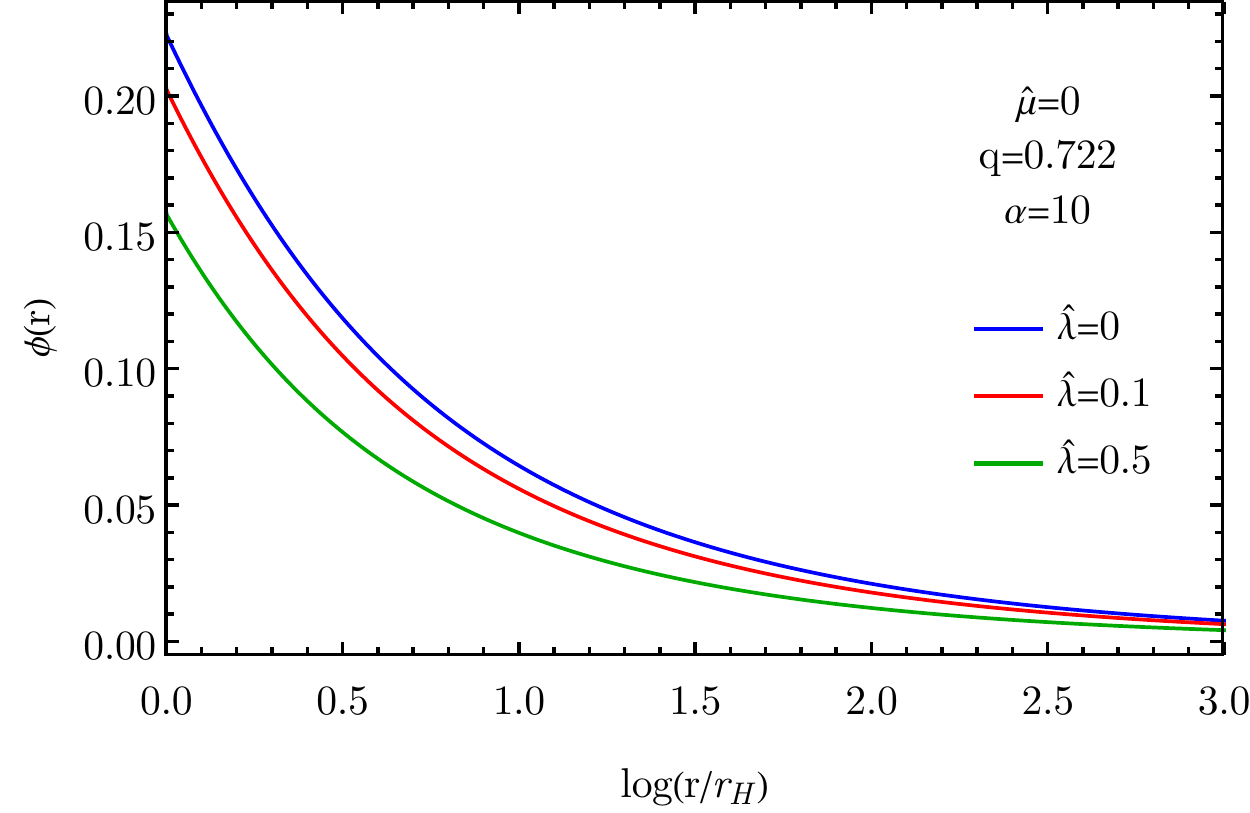}
\caption{Scalar field radial profiles for the scalarized case with several values of $\hat \lambda$ with $\hat \mu=0$.}
\label{fig:radprofiles2}
\end{figure}

 The radial profiles for the scalar field were also studied - \textit{c.f.} Fig. \ref{fig:radprofiles}. The mass term, as expected, has a suppressing effect on the scalar field (as seen, for instance, in the scalar field value at the event horizon). Also, the mass term leads to a scalar field radial profile more concentrated in the neighborhood of the event horizon and a (much) faster decay. This is expected since the decay is approximately exponential with the scalar field mass. The radial profiles of scalarized solutions with self-coupling $\lambda$ can be observed in Fig. \ref{fig:radprofiles2}. The self-coupling has a suppressing effect on the scalar field value at the horizon, leading to a scalar field profile that vanishes faster. On table \ref{tab:quantitiesradial} we present the values of some characteristic quantities of several scalarized solutions, such as the scalar charge $Q_s$ which is interesting from a phenomenological point of view as it would be associated with the dipolar scalar radiation emission in BH binaries \cite{Berti:2015itd,Yagi:2011xp,Berti:2018cxi}. We observe that both the scalar field mass and self-interaction parameter suppress the scalar charge value, with the former having a more predominant effect as, \textit{e.g.}, values of $\hat \mu \sim \mathcal{O}(10^{-2})$ can lead to almost negligible values of the scalar charge $\hat Q_s \sim \mathcal{O}(10^{-4})$.

\begin{table}[ht!]
    \centering
    \begin{tabular}{|c|c||c|c|c|c|c|c|c}\hline
        $\hat \mu$ & $\hat \lambda$ & $\hat Q_s$ & $a_H$ & $M$ & $-\Phi_e$ & $T_H$\\ \hline \hline
        0.0 & 0.0 & 0.343 & 0.726 & 1.388 & 0.366 & 0.029\\ \hline
        0.0 & 0.3 & 0.216 & 0.724 & 1.390 & 0.390 & 0.028 \\ \hline
        0.0 & 0.5 & 0.185 & 0.723 & 1.391 & 0.396 & 0.028\\ \hline
        0.01 & 0.0 & $\sim 10^{-3}$ & 0.724 & 1.389 & 0.375 & 0.029\\ \hline
        0.05 & 0.0 & $\sim 10^{-4}$ & 0.723 & 1.391 & 0.397 & 0.028\\ \hline
        0.01 & 0.3 & $\sim 10^{-3}$ & 0.722 & 1.391 & 0.392 & 0.028\\ \hline
        0.05 & 0.5 & $\sim 10^{-4}$ & 0.723 & 1.391 & 0.406 & 0.028\\ \hline
    \end{tabular}
    \caption{Characteristic quantities for several scalarized BH solutions with $q=0.72$ and $\alpha=10$. A big suppression on the scalar charge can be observed when there is a non-zero mass term.} 
    \label{tab:quantitiesradial}
\end{table}

\subsection{Effective potential for spherical perturbations and stability}
\par
\begin{figure}[ht!]
\centering
\includegraphics[width=.5\textwidth]{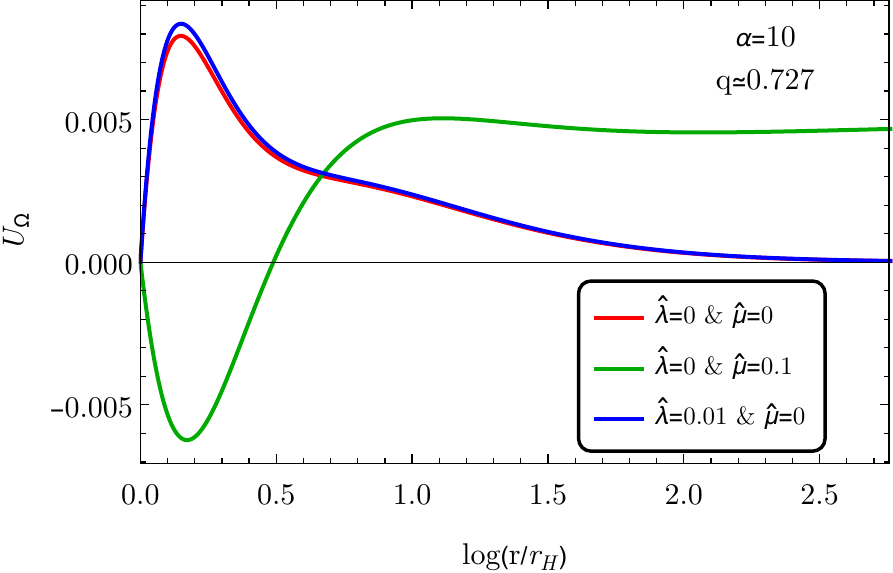}\hfill
\includegraphics[width=.5\textwidth]{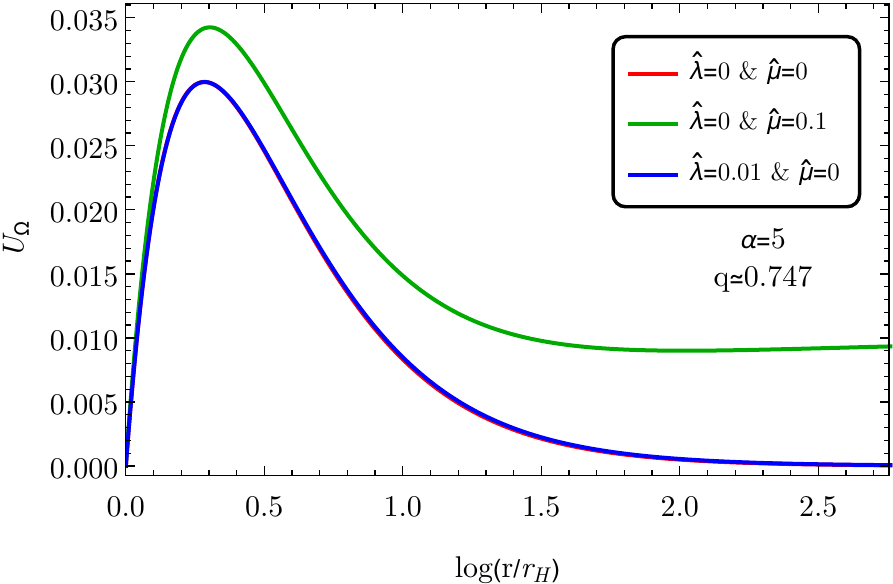}
\caption{(Left) Effective potential for several scalarized solutions. (Right) Effective potential for several dilaton solutions.}
\label{fig:effpot}
\end{figure}

 The effective potential for spherical perturbations was computed (\textit{c.f.} Fig. \ref{fig:effpot}) for both scalarized (left panel) and dilatonic (right panel) solutions. In the scalarized case the effective potential reveals that the self-interacting, massless solutions generically yield an everywhere positive effective potential with vanishing asymptotic values, thus being free of instabilities. On the the other hand solutions with a mass term that are close to the bifurcation point (\textit{i.e.}, whose charge to mass ratio is close to the existence line in the domain of existence) generically yield a negative region in the effective potential, thus instabilities cannot be excluded \textit{a priori}. In Ref. \cite{MassTermMyung} it was shown that these negative regions for the particular case $q=0.7$ ($n=0$) do not correspond to instabilities, suggesting that these solutions are perturbatively stable. Next we will generalize the results of Ref \cite{MassTermMyung} for a broader range of parameter space. An interesting feature of these massive solutions is the asymptotic value of the effective potential $U_\Omega \to \mu^2/2$ as $r\to \infty$ as can easily be observed in Eq. \eqref{eq:effectivepot}. For the dilatonic case all tackled solutions generically yield an everywhere positive effective potential with zero as the lowest of the asymptotic values, thus being free of instabilities.

\par
\begin{figure}[ht!]
\centering
\includegraphics[width=.5\textwidth]{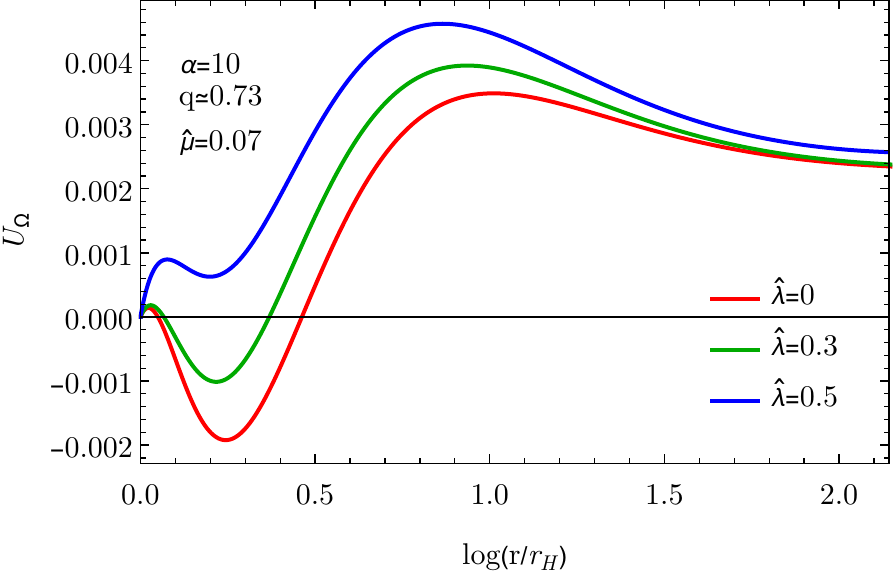}\hfill
\includegraphics[width=.5\textwidth]{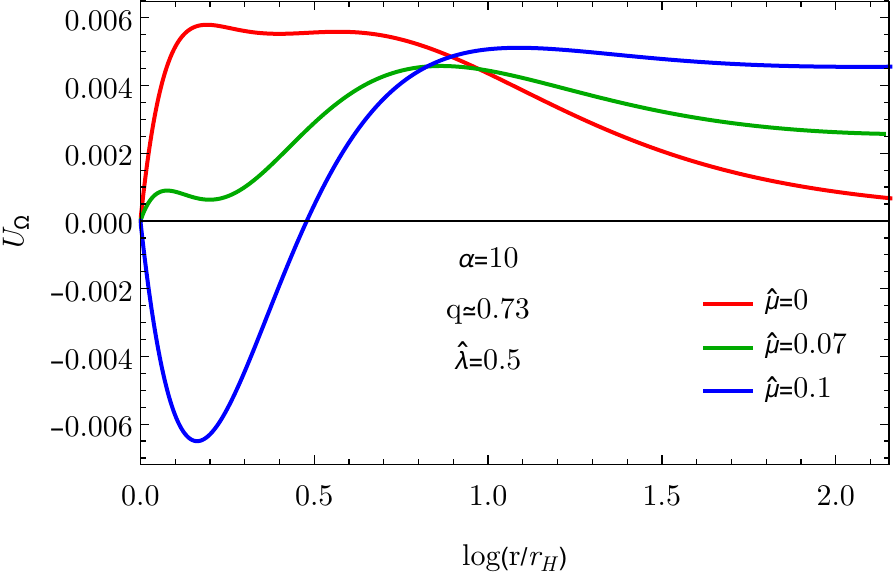}
\caption{Competition between the effects of $\hat \mu$ and $\hat \lambda$ for the effective potential of the scalarized solutions. (Left) Effective potential for several scalarized solutions keeping $\hat \mu=0.07$ fixed while varying $\hat \lambda$. (Right) Effective potential for several scalarized solutions keeping $\hat \lambda=0.5$ fixed while varying $\hat \mu$.}
\label{fig:effpot2}
\end{figure}

 To characterize the possible competition between $\hat \mu$ and $\hat \lambda$ on the effective potential, we obtained the profile of $U_\Omega$ for several scalarized solutions while fixing $\hat \mu$ and varying $\hat \lambda$ (Fig. \ref{fig:effpot2}, Left) and fixing $\hat \lambda$ while varying $\hat \mu$ (Fig. \ref{fig:effpot2}, Right). We observe that higher values of $\hat \lambda$ enforce stability such that for sufficiently high values of the self-coupling, the massive solutions yield an everywhere positive effective potential. On the other hand, higher values of the scalar field mass cause a depthening of the well of the effective potential.

\par
\begin{figure}[ht!]
\centering
\includegraphics[width=.5\textwidth]{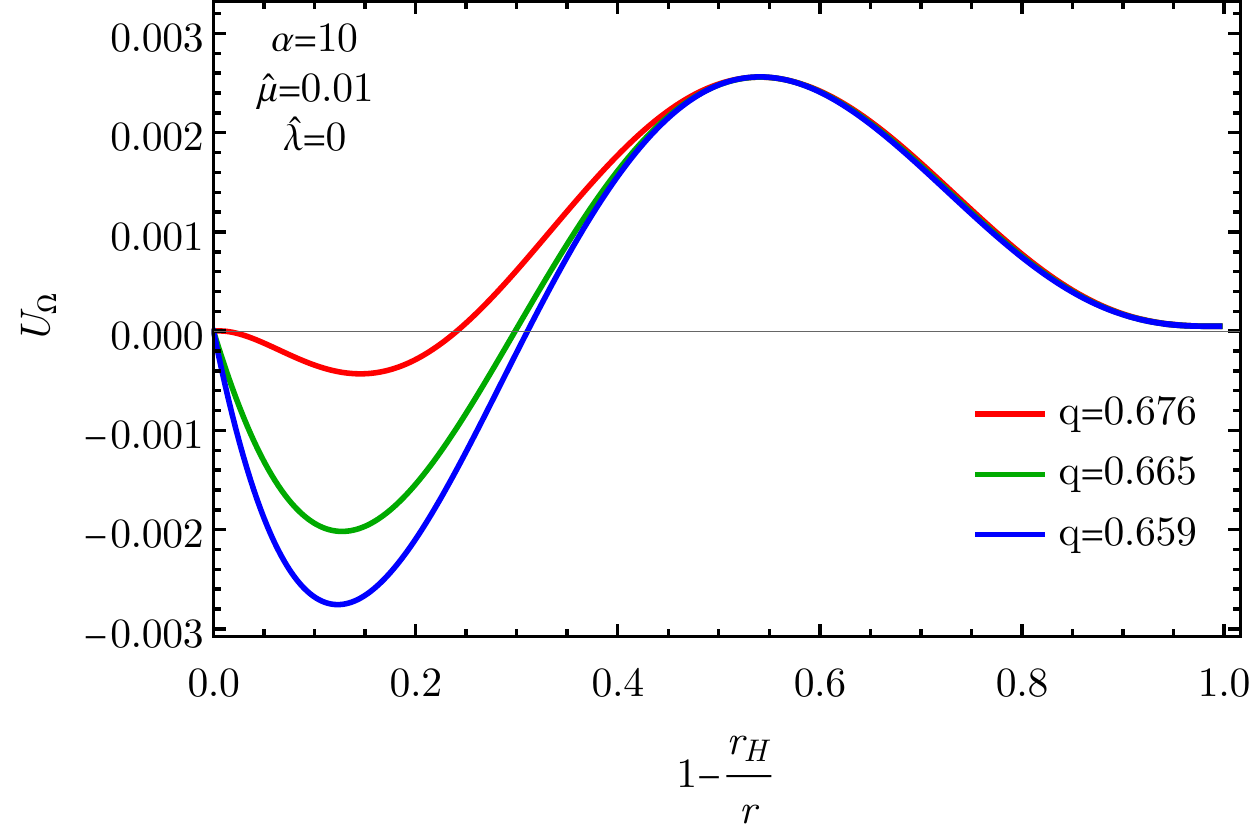}\hfill
\includegraphics[width=.5\textwidth]{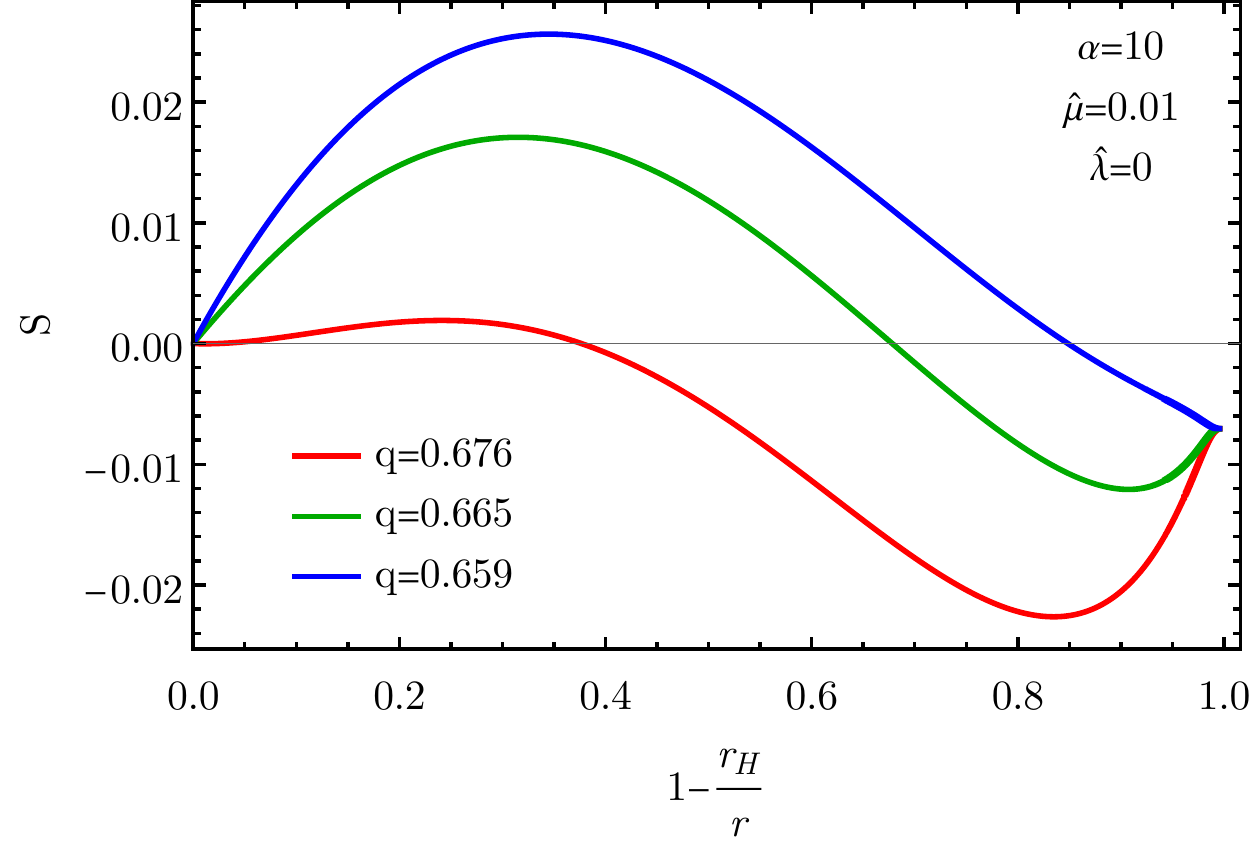}\vfill
\includegraphics[width=.5\textwidth]{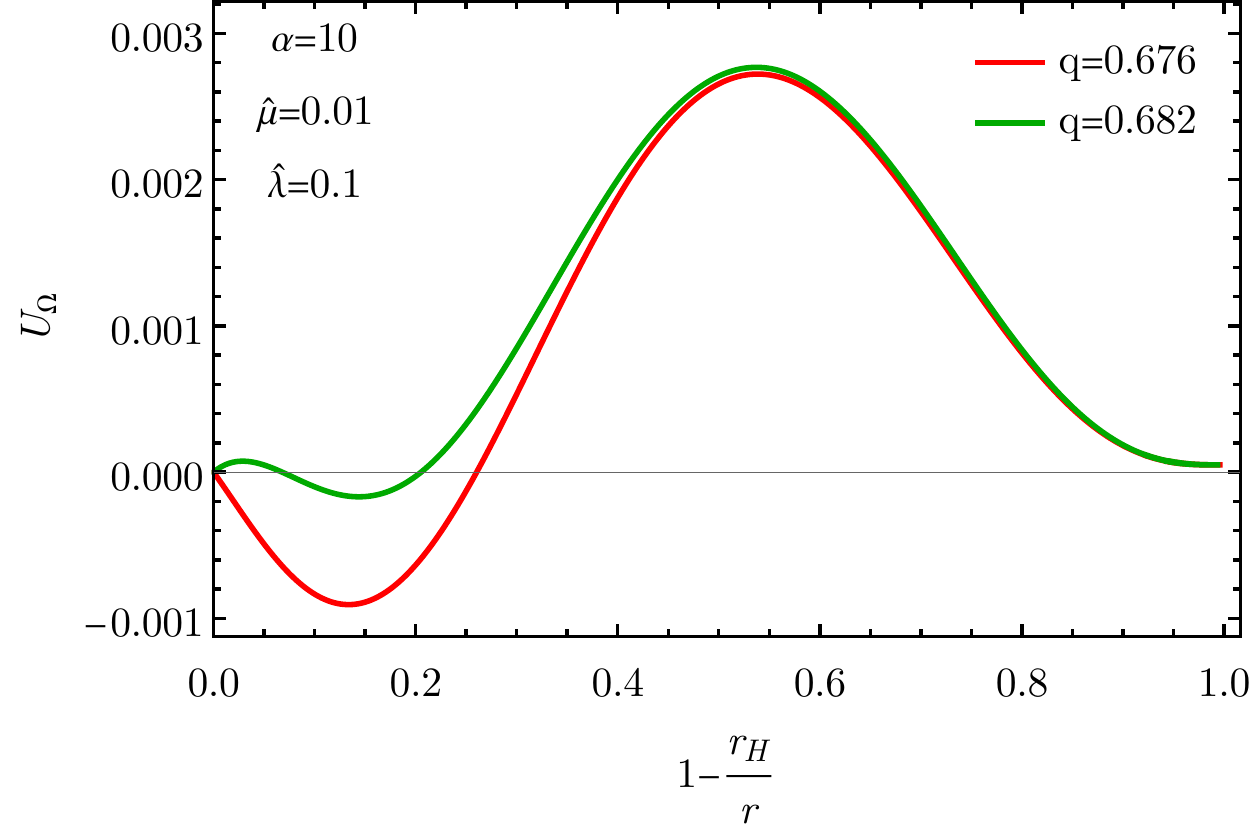}\hfill
\includegraphics[width=.5\textwidth]{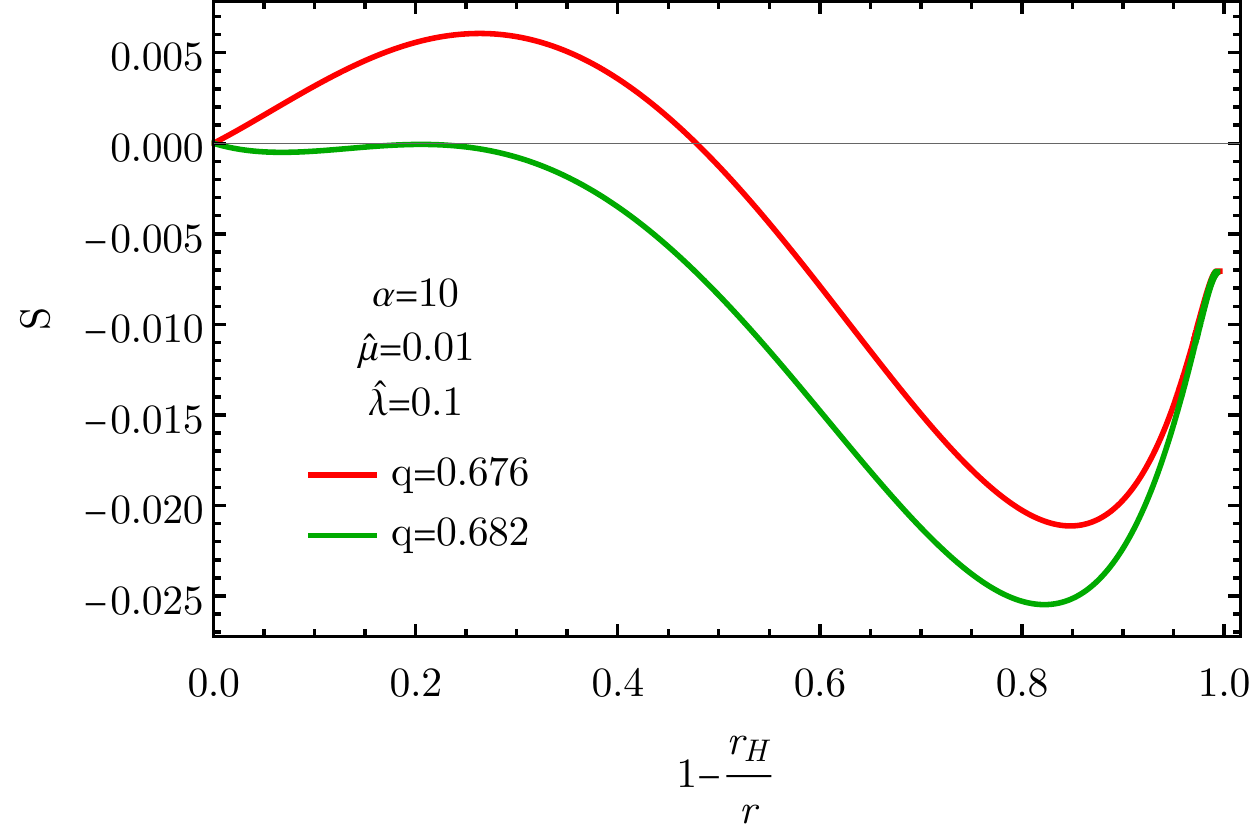}\vfill
\includegraphics[width=.5\textwidth]{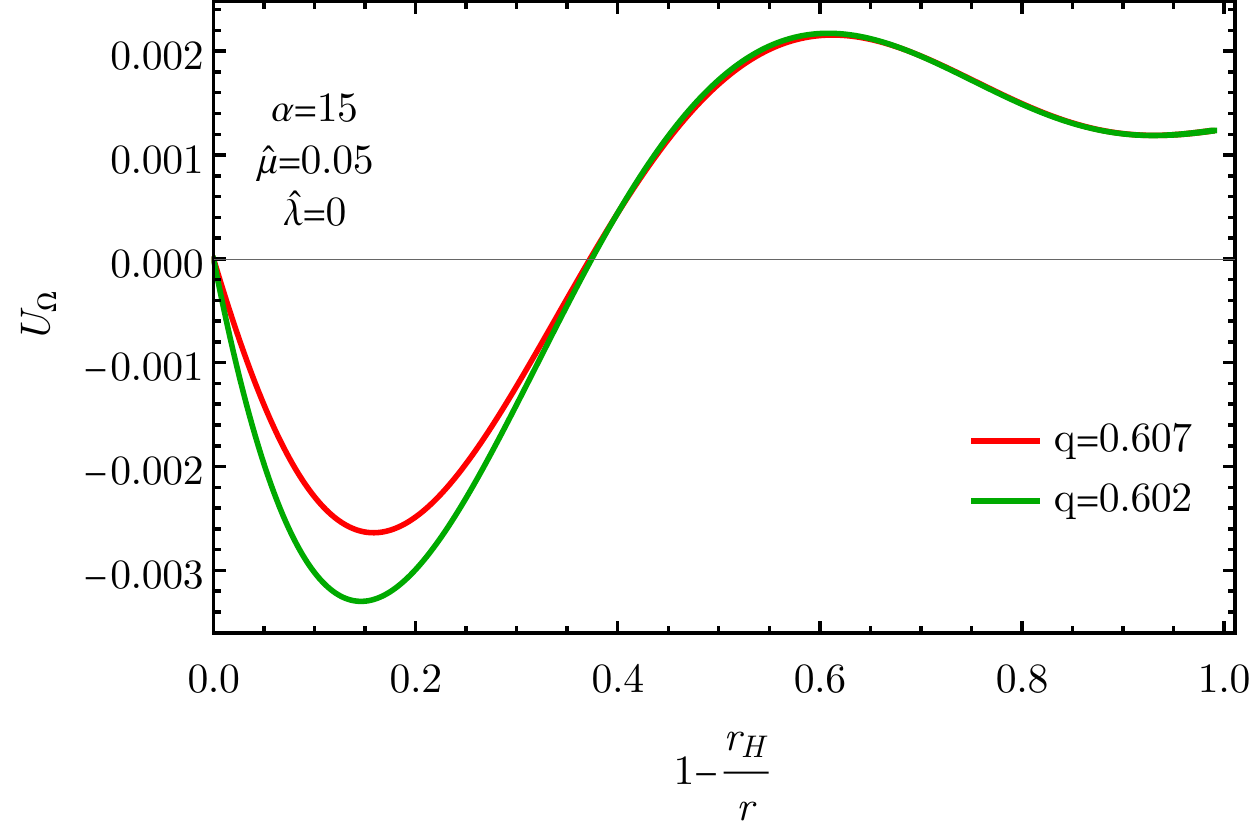}\hfill
\includegraphics[width=.5\textwidth]{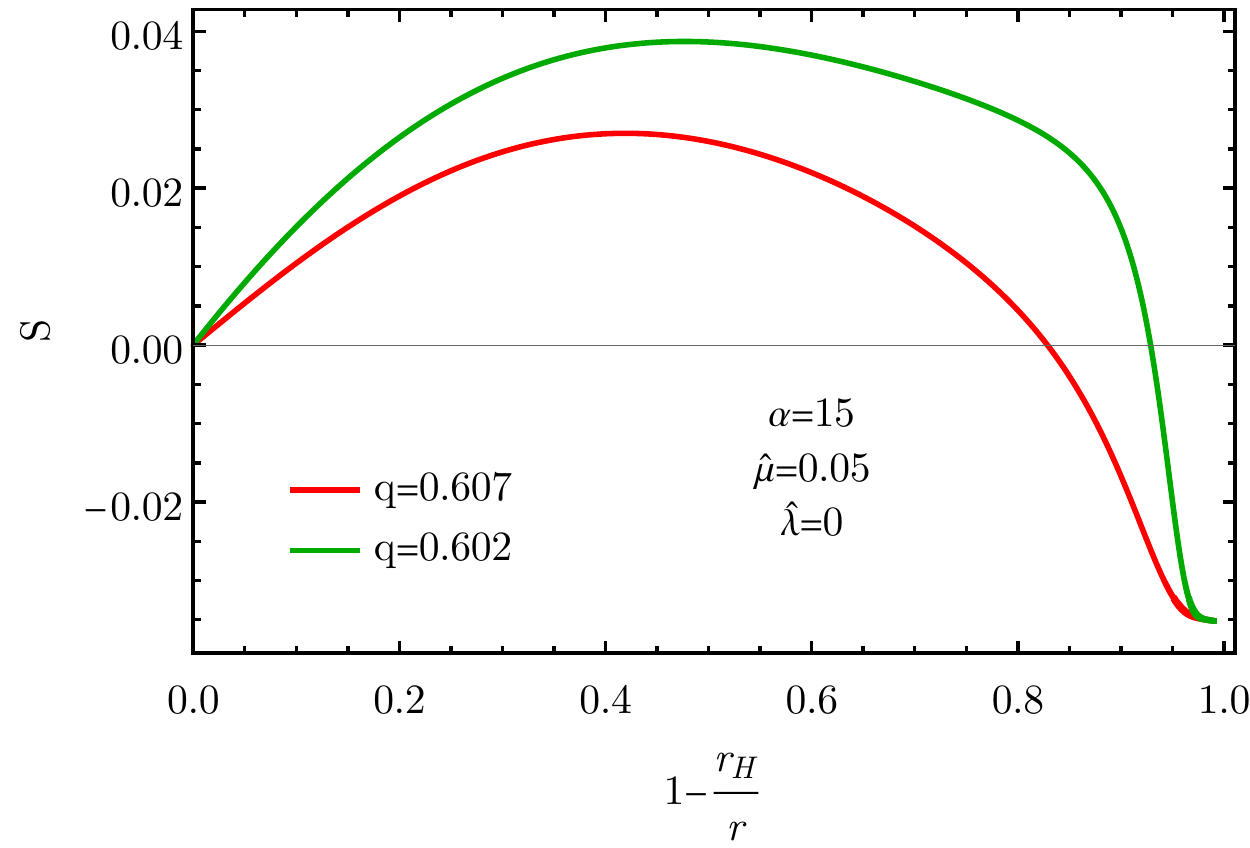}\vfill
\includegraphics[width=.5\textwidth]{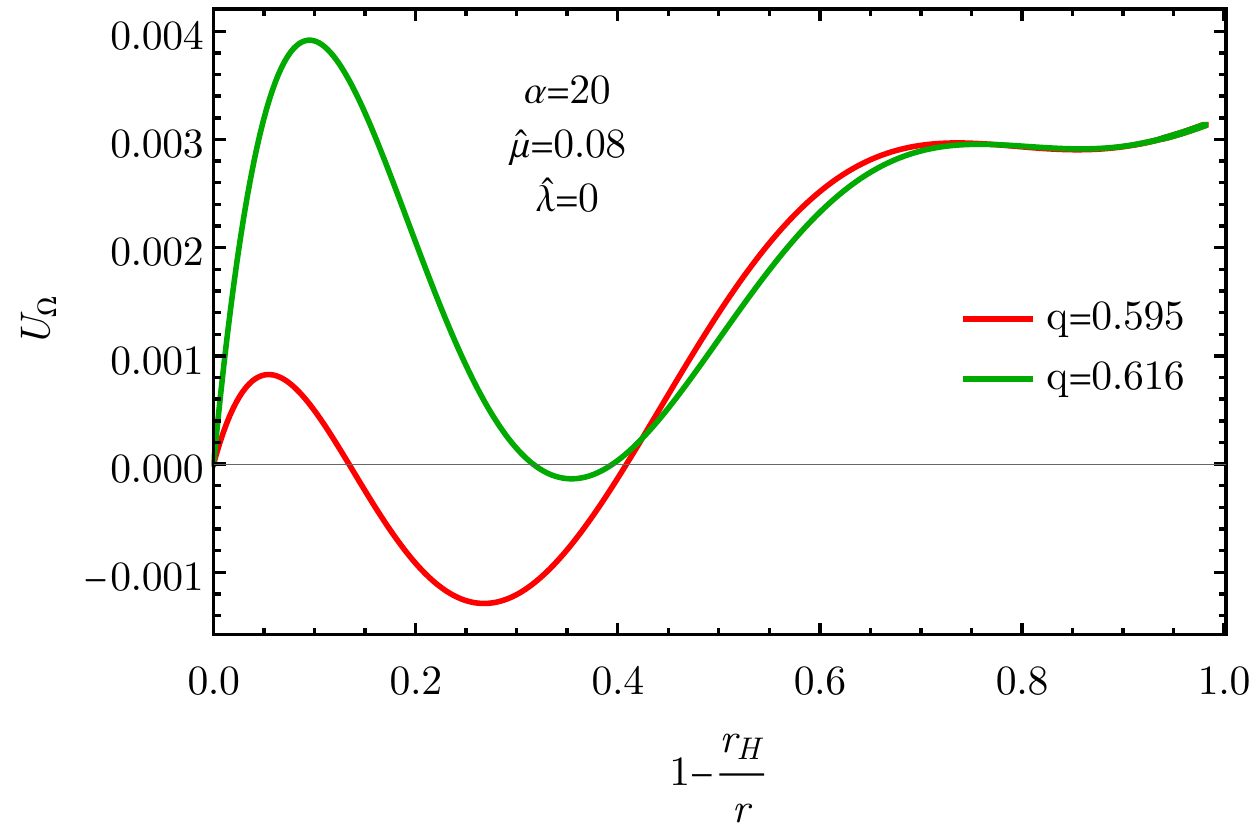}\hfill
\includegraphics[width=.5\textwidth]{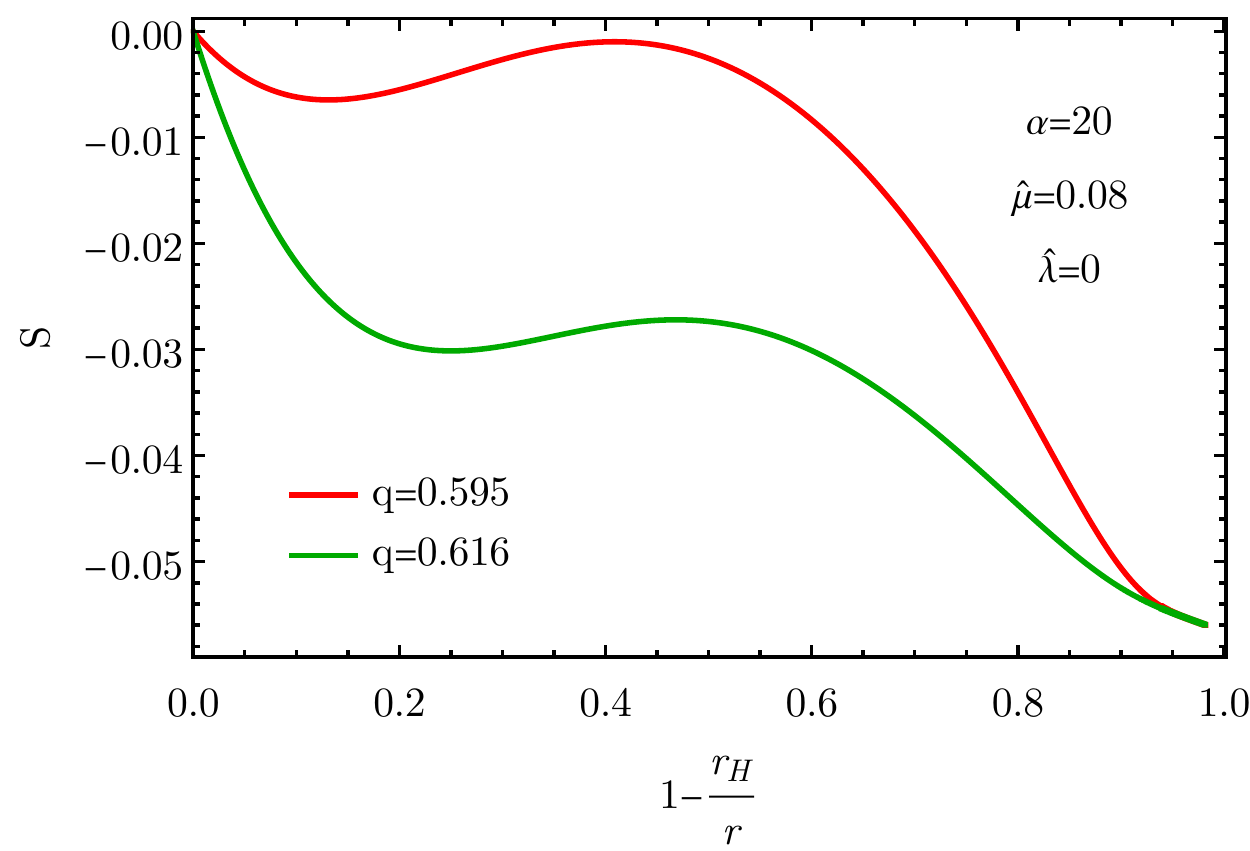}
\caption{A few examples of black hole solutions effective potential $U_\Omega$ that contain a negative well (left) and the respective deformation functions $S$ (right). It was always possible to obtain the deformation functions for all the tackled solutions, thus we conclude that the black holes are radially stable.}
\label{fig:sdeform}
\end{figure}

 Using the S-deformation method (\textit{c.f.} Sec. \ref{subsec:effpot}) we studied the radial stability of black hole solutions whose effective potential contains a negative well region. Resorting to the built-in function \textit{NDSolve} of the software \textit{Mathematica} we have numerically integrated Eq. \eqref{eq:sdef}, reading off the potential $U_\Omega$ from the numerical scalarized BH solutions. It was always possible to obtain a regular deformation function $S$ for all the studied solutions (\textit{c.f.} Fig. \ref{fig:sdeform} for a few examples), thus we conclude that the black hole solutions are radially stable. 

%
\section{Conclusions}\label{S4}
%
\par In this work we have studied the impact of scalar field mass and self-interactions on two paradigmatic EMS models - the dilaton and scalarized cases. Depending on the choice of the coupling $f(\phi)$, the model can accommodate BHs with scalar-hair and, may or may not accommodate the standard RN BH of electrovacuum. In the first case, the RN BH may become unstable against scalar perturbations and spontaneously develop scalar-hair (spontaneous scalarization).
\par It was found that the presence of a mass term alters the threshold for the onset of scalarization (which is independent of the self-interaction $\lambda$) as was already observed for eSTGB in Ref. \cite{Macedo:2019sem}. The results obtained for the bifurcation points of the $n=0$ mode of scalarized BHs agree with the ones obtained in Ref. \cite{MassTermMyung} for the specific case of $q=0.7$, and are presented in a scale-invariant form.
\par The domain of existence is bounded by critical lines where BH solutions are singular for both dilaton and scalarized BHs. The presence of a scalar field mass term narrows the domain of existence as the critical set occurs for smaller BH charge to mass ratios, while the scalar-field self-interaction has the opposite effect. For both couplings it was generically found that scalarized BHs are thermodynamically preferred over the scalar-free counterparts in the region of non-uniqueness ($q\leq 1$). We observed that the scalar field self-coupling $\lambda$ suppresses the scalar field value at the horizon, while the mass term, besides causing a suppression of the scalar field at the horizon, also causes an exponential decay. Both $\lambda$ and $\mu$ were found to suppress the values of the scalar charge. In what concerns the effective potential for spherical perturbations, for the dilaton coupling all solutions generically yield an everywhere positive effective potential with zero as the lowest of the asymptotic values, thus being free of instabilities. The same behavior is observed in the scalarized case for the massless but self-interacting case. In the massive scalar field case, the potential generically yields a negative region close to the existence line, that does not correspond to an instability as was shown via the S-deformation method. We characterized the competition between the scalar field mass and self-interactions on the effective potential, suggesting once again a stability enforcement by the self-coupling $\lambda$ and that the total effects of both a mass term and the self-coupling can be perceived as a combination of the single effects. For high enough self-coupling $\lambda$ it is always possible to find a black hole solution whose effective potential is everywhere positive, thus necessarily free of instabilities. Our analysis then suggests that the self-coupling $\lambda$ has a "stabilizing" effect on the BH solutions, mimicking the behaviour observed for eSTGB models in \cite{Macedo:2019sem}, completing a parallelism between the two models.
\par A closing remark: a preliminary study of dyonic solutions (electrically and magnetically charged BH solutions) was performed in a similar fashion as in section \ref{S3}, leading to similar global conclusions. However, as in the massless, non-self-interacting case the domain of existence is bounded by an extremal set, at which BHs are extremal. This extremal set is obtained, for the same $\alpha$, for smaller values of $q$ if a mass term is considered in spite of the massless case, and for higher values of $q$ if a non-zero self-coupling is considered (similarly to what is observed in the purely electric case for the critical set).

\section*{Acknowledgements}

P.F. is supported by the Royal Society grant RGF/EA/180022 and acknowledges support from  the  project CERN/FIS-PAR/0027/2019. The author would like to thank Alexandre M. Pombo for providing an initial version of the code used throughout this work and Carlos A. R. Herdeiro, Eugen Radu and Pedro Carrilho for useful comments on the manuscript.

\bibliography{biblio}

\begin{thebibliography}{52}%
\makeatletter
\providecommand \@ifxundefined [1]{%
 \@ifx{#1\undefined}
}%
\providecommand \@ifnum [1]{%
 \ifnum #1\expandafter \@firstoftwo
 \else \expandafter \@secondoftwo
 \fi
}%
\providecommand \@ifx [1]{%
 \ifx #1\expandafter \@firstoftwo
 \else \expandafter \@secondoftwo
 \fi
}%
\providecommand \natexlab [1]{#1}%
\providecommand \enquote  [1]{``#1''}%
\providecommand \bibnamefont  [1]{#1}%
\providecommand \bibfnamefont [1]{#1}%
\providecommand \citenamefont [1]{#1}%
\providecommand \href@noop [0]{\@secondoftwo}%
\providecommand \href [0]{\begingroup \@sanitize@url \@href}%
\providecommand \@href[1]{\@@startlink{#1}\@@href}%
\providecommand \@@href[1]{\endgroup#1\@@endlink}%
\providecommand \@sanitize@url [0]{\catcode `\\12\catcode `\$12\catcode
  `\&12\catcode `\#12\catcode `\^12\catcode `\_12\catcode `\%12\relax}%
\providecommand \@@startlink[1]{}%
\providecommand \@@endlink[0]{}%
\providecommand \url  [0]{\begingroup\@sanitize@url \@url }%
\providecommand \@url [1]{\endgroup\@href {#1}{\urlprefix }}%
\providecommand \urlprefix  [0]{URL }%
\providecommand \Eprint [0]{\href }%
\providecommand \doibase [0]{http://dx.doi.org/}%
\providecommand \selectlanguage [0]{\@gobble}%
\providecommand \bibinfo  [0]{\@secondoftwo}%
\providecommand \bibfield  [0]{\@secondoftwo}%
\providecommand \translation [1]{[#1]}%
\providecommand \BibitemOpen [0]{}%
\providecommand \bibitemStop [0]{}%
\providecommand \bibitemNoStop [0]{.\EOS\space}%
\providecommand \EOS [0]{\spacefactor3000\relax}%
\providecommand \BibitemShut  [1]{\csname bibitem#1\endcsname}%
\let\auto@bib@innerbib\@empty
\bibitem [{\citenamefont {Garfinkle}\ \emph {et~al.}(1991)\citenamefont
  {Garfinkle}, \citenamefont {Horowitz},\ and\ \citenamefont
  {Strominger}}]{DilatonGarfinkle}%
  \BibitemOpen
  \bibfield  {author} {\bibinfo {author} {\bibfnamefont {D.}~\bibnamefont
  {Garfinkle}}, \bibinfo {author} {\bibfnamefont {G.~T.}\ \bibnamefont
  {Horowitz}}, \ and\ \bibinfo {author} {\bibfnamefont {A.}~\bibnamefont
  {Strominger}},\ }\href {\doibase 10.1103/PhysRevD.43.3140} {\bibfield
  {journal} {\bibinfo  {journal} {Phys. Rev. D}\ }\textbf {\bibinfo {volume}
  {43}},\ \bibinfo {pages} {3140} (\bibinfo {year} {1991})}\BibitemShut
  {NoStop}%
\bibitem [{\citenamefont {Gibbons}\ and\ \citenamefont {ichi
  Maeda}(1988)}]{DilatonGibbons}%
  \BibitemOpen
  \bibfield  {author} {\bibinfo {author} {\bibfnamefont {G.}~\bibnamefont
  {Gibbons}}\ and\ \bibinfo {author} {\bibfnamefont {K.}~\bibnamefont {ichi
  Maeda}},\ }\href {\doibase https://doi.org/10.1016/0550-3213(88)90006-5}
  {\bibfield  {journal} {\bibinfo  {journal} {Nuclear Physics B}\ }\textbf
  {\bibinfo {volume} {298}},\ \bibinfo {pages} {741 } (\bibinfo {year}
  {1988})}\BibitemShut {NoStop}%
\bibitem [{\citenamefont {Herdeiro}\ \emph {et~al.}(2018)\citenamefont
  {Herdeiro}, \citenamefont {Radu}, \citenamefont {Sanchis-Gual},\ and\
  \citenamefont {Font}}]{MainPaper1}%
  \BibitemOpen
  \bibfield  {author} {\bibinfo {author} {\bibfnamefont {C.~A.~R.}\
  \bibnamefont {Herdeiro}}, \bibinfo {author} {\bibfnamefont {E.}~\bibnamefont
  {Radu}}, \bibinfo {author} {\bibfnamefont {N.}~\bibnamefont {Sanchis-Gual}},
  \ and\ \bibinfo {author} {\bibfnamefont {J.~A.}\ \bibnamefont {Font}},\
  }\href {\doibase 10.1103/PhysRevLett.121.101102} {\bibfield  {journal}
  {\bibinfo  {journal} {Phys. Rev. Lett.}\ }\textbf {\bibinfo {volume} {121}},\
  \bibinfo {pages} {101102} (\bibinfo {year} {2018})},\ \Eprint
  {http://arxiv.org/abs/1806.05190} {arXiv:1806.05190 [gr-qc]} \BibitemShut
  {NoStop}%
\bibitem [{\citenamefont {Fernandes}\ \emph
  {et~al.}(2019{\natexlab{a}})\citenamefont {Fernandes}, \citenamefont
  {Herdeiro}, \citenamefont {Pombo}, \citenamefont {Radu},\ and\ \citenamefont
  {Sanchis-Gual}}]{MainPaper2}%
  \BibitemOpen
  \bibfield  {author} {\bibinfo {author} {\bibfnamefont {P.~G.~S.}\
  \bibnamefont {Fernandes}}, \bibinfo {author} {\bibfnamefont {C.~A.~R.}\
  \bibnamefont {Herdeiro}}, \bibinfo {author} {\bibfnamefont {A.~M.}\
  \bibnamefont {Pombo}}, \bibinfo {author} {\bibfnamefont {E.}~\bibnamefont
  {Radu}}, \ and\ \bibinfo {author} {\bibfnamefont {N.}~\bibnamefont
  {Sanchis-Gual}},\ }\href {\doibase 10.1088/1361-6382/ab23a1} {\bibfield
  {journal} {\bibinfo  {journal} {Class. Quant. Grav.}\ }\textbf {\bibinfo
  {volume} {36}},\ \bibinfo {pages} {134002} (\bibinfo {year}
  {2019}{\natexlab{a}})},\ \Eprint {http://arxiv.org/abs/1902.05079}
  {arXiv:1902.05079 [gr-qc]} \BibitemShut {NoStop}%
\bibitem [{\citenamefont {Astefanesei}\ \emph {et~al.}(2019)\citenamefont
  {Astefanesei}, \citenamefont {Herdeiro}, \citenamefont {Pombo},\ and\
  \citenamefont {Radu}}]{MainPaper3}%
  \BibitemOpen
  \bibfield  {author} {\bibinfo {author} {\bibfnamefont {D.}~\bibnamefont
  {Astefanesei}}, \bibinfo {author} {\bibfnamefont {C.}~\bibnamefont
  {Herdeiro}}, \bibinfo {author} {\bibfnamefont {A.}~\bibnamefont {Pombo}}, \
  and\ \bibinfo {author} {\bibfnamefont {E.}~\bibnamefont {Radu}},\ }\href
  {\doibase 10.1007/JHEP10(2019)078} {\bibfield  {journal} {\bibinfo  {journal}
  {JHEP}\ }\textbf {\bibinfo {volume} {10}},\ \bibinfo {pages} {078} (\bibinfo
  {year} {2019})},\ \Eprint {http://arxiv.org/abs/1905.08304} {arXiv:1905.08304
  [hep-th]} \BibitemShut {NoStop}%
\bibitem [{\citenamefont {Fernandes}\ \emph
  {et~al.}(2019{\natexlab{b}})\citenamefont {Fernandes}, \citenamefont
  {Herdeiro}, \citenamefont {Pombo}, \citenamefont {Radu},\ and\ \citenamefont
  {Sanchis-Gual}}]{MainPaper4}%
  \BibitemOpen
  \bibfield  {author} {\bibinfo {author} {\bibfnamefont {P.~G.~S.}\
  \bibnamefont {Fernandes}}, \bibinfo {author} {\bibfnamefont {C.~A.~R.}\
  \bibnamefont {Herdeiro}}, \bibinfo {author} {\bibfnamefont {A.~M.}\
  \bibnamefont {Pombo}}, \bibinfo {author} {\bibfnamefont {E.}~\bibnamefont
  {Radu}}, \ and\ \bibinfo {author} {\bibfnamefont {N.}~\bibnamefont
  {Sanchis-Gual}},\ }\href {\doibase 10.1103/PhysRevD.100.084045} {\bibfield
  {journal} {\bibinfo  {journal} {Phys. Rev.}\ }\textbf {\bibinfo {volume}
  {D100}},\ \bibinfo {pages} {084045} (\bibinfo {year} {2019}{\natexlab{b}})},\
  \Eprint {http://arxiv.org/abs/1908.00037} {arXiv:1908.00037 [gr-qc]}
  \BibitemShut {NoStop}%
\bibitem [{\citenamefont {Doneva}\ \emph {et~al.}(2010)\citenamefont {Doneva},
  \citenamefont {Yazadjiev}, \citenamefont {Kokkotas},\ and\ \citenamefont
  {Stefanov}}]{Doneva:2010ke}%
  \BibitemOpen
  \bibfield  {author} {\bibinfo {author} {\bibfnamefont {D.~D.}\ \bibnamefont
  {Doneva}}, \bibinfo {author} {\bibfnamefont {S.~S.}\ \bibnamefont
  {Yazadjiev}}, \bibinfo {author} {\bibfnamefont {K.~D.}\ \bibnamefont
  {Kokkotas}}, \ and\ \bibinfo {author} {\bibfnamefont {I.~Z.}\ \bibnamefont
  {Stefanov}},\ }\href {\doibase 10.1103/PhysRevD.82.064030} {\bibfield
  {journal} {\bibinfo  {journal} {Phys. Rev.}\ }\textbf {\bibinfo {volume}
  {D82}},\ \bibinfo {pages} {064030} (\bibinfo {year} {2010})},\ \Eprint
  {http://arxiv.org/abs/1007.1767} {arXiv:1007.1767 [gr-qc]} \BibitemShut
  {NoStop}%
\bibitem [{\citenamefont {Stefanov}\ \emph {et~al.}(2008)\citenamefont
  {Stefanov}, \citenamefont {Yazadjiev},\ and\ \citenamefont
  {Todorov}}]{Stefanov:2007eq}%
  \BibitemOpen
  \bibfield  {author} {\bibinfo {author} {\bibfnamefont {I.~Z.}\ \bibnamefont
  {Stefanov}}, \bibinfo {author} {\bibfnamefont {S.~S.}\ \bibnamefont
  {Yazadjiev}}, \ and\ \bibinfo {author} {\bibfnamefont {M.~D.}\ \bibnamefont
  {Todorov}},\ }\href {\doibase 10.1142/S0217732308028351} {\bibfield
  {journal} {\bibinfo  {journal} {Mod. Phys. Lett.}\ }\textbf {\bibinfo
  {volume} {A23}},\ \bibinfo {pages} {2915} (\bibinfo {year} {2008})},\ \Eprint
  {http://arxiv.org/abs/0708.4141} {arXiv:0708.4141 [gr-qc]} \BibitemShut
  {NoStop}%
\bibitem [{\citenamefont {Gubser}(2005)}]{Gubser:2005ih}%
  \BibitemOpen
  \bibfield  {author} {\bibinfo {author} {\bibfnamefont {S.~S.}\ \bibnamefont
  {Gubser}},\ }\href {\doibase 10.1088/0264-9381/22/23/013} {\bibfield
  {journal} {\bibinfo  {journal} {Class. Quant. Grav.}\ }\textbf {\bibinfo
  {volume} {22}},\ \bibinfo {pages} {5121} (\bibinfo {year} {2005})},\ \Eprint
  {http://arxiv.org/abs/hep-th/0505189} {arXiv:hep-th/0505189 [hep-th]}
  \BibitemShut {NoStop}%
\bibitem [{\citenamefont {Macedo}\ \emph {et~al.}(2019)\citenamefont {Macedo},
  \citenamefont {Sakstein}, \citenamefont {Berti}, \citenamefont {Gualtieri},
  \citenamefont {Silva},\ and\ \citenamefont {Sotiriou}}]{Macedo:2019sem}%
  \BibitemOpen
  \bibfield  {author} {\bibinfo {author} {\bibfnamefont {C.~F.~B.}\
  \bibnamefont {Macedo}}, \bibinfo {author} {\bibfnamefont {J.}~\bibnamefont
  {Sakstein}}, \bibinfo {author} {\bibfnamefont {E.}~\bibnamefont {Berti}},
  \bibinfo {author} {\bibfnamefont {L.}~\bibnamefont {Gualtieri}}, \bibinfo
  {author} {\bibfnamefont {H.~O.}\ \bibnamefont {Silva}}, \ and\ \bibinfo
  {author} {\bibfnamefont {T.~P.}\ \bibnamefont {Sotiriou}},\ }\href {\doibase
  10.1103/PhysRevD.99.104041} {\bibfield  {journal} {\bibinfo  {journal} {Phys.
  Rev.}\ }\textbf {\bibinfo {volume} {D99}},\ \bibinfo {pages} {104041}
  (\bibinfo {year} {2019})},\ \Eprint {http://arxiv.org/abs/1903.06784}
  {arXiv:1903.06784 [gr-qc]} \BibitemShut {NoStop}%
\bibitem [{\citenamefont {Doneva}\ \emph {et~al.}(2019)\citenamefont {Doneva},
  \citenamefont {Staykov},\ and\ \citenamefont {Yazadjiev}}]{Doneva:2019vuh}%
  \BibitemOpen
  \bibfield  {author} {\bibinfo {author} {\bibfnamefont {D.~D.}\ \bibnamefont
  {Doneva}}, \bibinfo {author} {\bibfnamefont {K.~V.}\ \bibnamefont {Staykov}},
  \ and\ \bibinfo {author} {\bibfnamefont {S.~S.}\ \bibnamefont {Yazadjiev}},\
  }\href {\doibase 10.1103/PhysRevD.99.104045} {\bibfield  {journal} {\bibinfo
  {journal} {Phys. Rev.}\ }\textbf {\bibinfo {volume} {D99}},\ \bibinfo {pages}
  {104045} (\bibinfo {year} {2019})},\ \Eprint
  {http://arxiv.org/abs/1903.08119} {arXiv:1903.08119 [gr-qc]} \BibitemShut
  {NoStop}%
\bibitem [{\citenamefont {Silva}\ \emph {et~al.}(2018)\citenamefont {Silva},
  \citenamefont {Sakstein}, \citenamefont {Gualtieri}, \citenamefont
  {Sotiriou},\ and\ \citenamefont {Berti}}]{Silva_2018}%
  \BibitemOpen
  \bibfield  {author} {\bibinfo {author} {\bibfnamefont {H.~O.}\ \bibnamefont
  {Silva}}, \bibinfo {author} {\bibfnamefont {J.}~\bibnamefont {Sakstein}},
  \bibinfo {author} {\bibfnamefont {L.}~\bibnamefont {Gualtieri}}, \bibinfo
  {author} {\bibfnamefont {T.~P.}\ \bibnamefont {Sotiriou}}, \ and\ \bibinfo
  {author} {\bibfnamefont {E.}~\bibnamefont {Berti}},\ }\href {\doibase
  10.1103/physrevlett.120.131104} {\bibfield  {journal} {\bibinfo  {journal}
  {Physical Review Letters}\ }\textbf {\bibinfo {volume} {120}} (\bibinfo
  {year} {2018}),\ 10.1103/physrevlett.120.131104}\BibitemShut {NoStop}%
\bibitem [{\citenamefont {Doneva}\ and\ \citenamefont
  {Yazadjiev}(2018)}]{Doneva_2018}%
  \BibitemOpen
  \bibfield  {author} {\bibinfo {author} {\bibfnamefont {D.~D.}\ \bibnamefont
  {Doneva}}\ and\ \bibinfo {author} {\bibfnamefont {S.~S.}\ \bibnamefont
  {Yazadjiev}},\ }\href {\doibase 10.1103/physrevlett.120.131103} {\bibfield
  {journal} {\bibinfo  {journal} {Physical Review Letters}\ }\textbf {\bibinfo
  {volume} {120}} (\bibinfo {year} {2018}),\
  10.1103/physrevlett.120.131103}\BibitemShut {NoStop}%
\bibitem [{\citenamefont {Antoniou}\ \emph {et~al.}(2018)\citenamefont
  {Antoniou}, \citenamefont {Bakopoulos},\ and\ \citenamefont
  {Kanti}}]{Antoniou_2018}%
  \BibitemOpen
  \bibfield  {author} {\bibinfo {author} {\bibfnamefont {G.}~\bibnamefont
  {Antoniou}}, \bibinfo {author} {\bibfnamefont {A.}~\bibnamefont
  {Bakopoulos}}, \ and\ \bibinfo {author} {\bibfnamefont {P.}~\bibnamefont
  {Kanti}},\ }\href {\doibase 10.1103/physrevlett.120.131102} {\bibfield
  {journal} {\bibinfo  {journal} {Physical Review Letters}\ }\textbf {\bibinfo
  {volume} {120}} (\bibinfo {year} {2018}),\
  10.1103/physrevlett.120.131102}\BibitemShut {NoStop}%
\bibitem [{\citenamefont {Cunha}\ \emph {et~al.}(2019)\citenamefont {Cunha},
  \citenamefont {Herdeiro},\ and\ \citenamefont {Radu}}]{Cunha:2019dwb}%
  \BibitemOpen
  \bibfield  {author} {\bibinfo {author} {\bibfnamefont {P.~V.~P.}\
  \bibnamefont {Cunha}}, \bibinfo {author} {\bibfnamefont {C.~A.~R.}\
  \bibnamefont {Herdeiro}}, \ and\ \bibinfo {author} {\bibfnamefont
  {E.}~\bibnamefont {Radu}},\ }\href {\doibase 10.1103/PhysRevLett.123.011101}
  {\bibfield  {journal} {\bibinfo  {journal} {Phys. Rev. Lett.}\ }\textbf
  {\bibinfo {volume} {123}},\ \bibinfo {pages} {011101} (\bibinfo {year}
  {2019})},\ \Eprint {http://arxiv.org/abs/1904.09997} {arXiv:1904.09997
  [gr-qc]} \BibitemShut {NoStop}%
\bibitem [{\citenamefont {Zaja\v~cek}\ and\ \citenamefont
  {Tursunov}(2019)}]{Zajacek:2019kla}%
  \BibitemOpen
  \bibfield  {author} {\bibinfo {author} {\bibfnamefont {M.}~\bibnamefont
  {Zaja\v~cek}}\ and\ \bibinfo {author} {\bibfnamefont {A.}~\bibnamefont
  {Tursunov}},\ }\href@noop {} {\  (\bibinfo {year} {2019})},\ \Eprint
  {http://arxiv.org/abs/1904.04654} {arXiv:1904.04654 [astro-ph.GA]}
  \BibitemShut {NoStop}%
\bibitem [{\citenamefont {Gibbons}(1975)}]{Gibbons:1975kk}%
  \BibitemOpen
  \bibfield  {author} {\bibinfo {author} {\bibfnamefont {G.}~\bibnamefont
  {Gibbons}},\ }\href {\doibase 10.1007/BF01609829} {\bibfield  {journal}
  {\bibinfo  {journal} {Commun. Math. Phys.}\ }\textbf {\bibinfo {volume}
  {44}},\ \bibinfo {pages} {245} (\bibinfo {year} {1975})}\BibitemShut
  {NoStop}%
\bibitem [{\citenamefont {Cardoso}\ \emph {et~al.}(2016)\citenamefont
  {Cardoso}, \citenamefont {Macedo}, \citenamefont {Pani},\ and\ \citenamefont
  {Ferrari}}]{Cardoso:2016olt}%
  \BibitemOpen
  \bibfield  {author} {\bibinfo {author} {\bibfnamefont {V.}~\bibnamefont
  {Cardoso}}, \bibinfo {author} {\bibfnamefont {C.~F.~B.}\ \bibnamefont
  {Macedo}}, \bibinfo {author} {\bibfnamefont {P.}~\bibnamefont {Pani}}, \ and\
  \bibinfo {author} {\bibfnamefont {V.}~\bibnamefont {Ferrari}},\ }\href
  {\doibase 10.1088/1475-7516/2016/05/054} {\bibfield  {journal} {\bibinfo
  {journal} {JCAP}\ }\textbf {\bibinfo {volume} {05}},\ \bibinfo {pages} {054}
  (\bibinfo {year} {2016})},\ \bibinfo {note} {[Erratum: JCAP 04, E01
  (2020)]},\ \Eprint {http://arxiv.org/abs/1604.07845} {arXiv:1604.07845
  [hep-ph]} \BibitemShut {NoStop}%
\bibitem [{\citenamefont {De~Rujula}\ \emph {et~al.}(1990)\citenamefont
  {De~Rujula}, \citenamefont {Glashow},\ and\ \citenamefont
  {Sarid}}]{DeRujula:1989fe}%
  \BibitemOpen
  \bibfield  {author} {\bibinfo {author} {\bibfnamefont {A.}~\bibnamefont
  {De~Rujula}}, \bibinfo {author} {\bibfnamefont {S.}~\bibnamefont {Glashow}},
  \ and\ \bibinfo {author} {\bibfnamefont {U.}~\bibnamefont {Sarid}},\ }\href
  {\doibase 10.1016/0550-3213(90)90227-5} {\bibfield  {journal} {\bibinfo
  {journal} {Nucl. Phys. B}\ }\textbf {\bibinfo {volume} {333}},\ \bibinfo
  {pages} {173} (\bibinfo {year} {1990})}\BibitemShut {NoStop}%
\bibitem [{\citenamefont {Plestid}\ \emph {et~al.}(2020)\citenamefont
  {Plestid}, \citenamefont {Takhistov}, \citenamefont {Tsai}, \citenamefont
  {Bringmann}, \citenamefont {Kusenko},\ and\ \citenamefont
  {Pospelov}}]{Plestid:2020kdm}%
  \BibitemOpen
  \bibfield  {author} {\bibinfo {author} {\bibfnamefont {R.}~\bibnamefont
  {Plestid}}, \bibinfo {author} {\bibfnamefont {V.}~\bibnamefont {Takhistov}},
  \bibinfo {author} {\bibfnamefont {Y.-D.}\ \bibnamefont {Tsai}}, \bibinfo
  {author} {\bibfnamefont {T.}~\bibnamefont {Bringmann}}, \bibinfo {author}
  {\bibfnamefont {A.}~\bibnamefont {Kusenko}}, \ and\ \bibinfo {author}
  {\bibfnamefont {M.}~\bibnamefont {Pospelov}},\ }\href@noop {} {\  (\bibinfo
  {year} {2020})},\ \Eprint {http://arxiv.org/abs/2002.11732} {arXiv:2002.11732
  [hep-ph]} \BibitemShut {NoStop}%
\bibitem [{\citenamefont {Ackerman}\ \emph {et~al.}(2008)\citenamefont
  {Ackerman}, \citenamefont {Buckley}, \citenamefont {Carroll},\ and\
  \citenamefont {Kamionkowski}}]{Ackerman:mha}%
  \BibitemOpen
  \bibfield  {author} {\bibinfo {author} {\bibfnamefont {L.}~\bibnamefont
  {Ackerman}}, \bibinfo {author} {\bibfnamefont {M.~R.}\ \bibnamefont
  {Buckley}}, \bibinfo {author} {\bibfnamefont {S.~M.}\ \bibnamefont
  {Carroll}}, \ and\ \bibinfo {author} {\bibfnamefont {M.}~\bibnamefont
  {Kamionkowski}},\ }\href {\doibase 10.1103/PhysRevD.79.023519} {\ ,\ \bibinfo
  {pages} {277} (\bibinfo {year} {2008})},\ \Eprint
  {http://arxiv.org/abs/0810.5126} {arXiv:0810.5126 [hep-ph]} \BibitemShut
  {NoStop}%
\bibitem [{\citenamefont {Abbott}\ \emph {et~al.}(2016)\citenamefont {Abbott}
  \emph {et~al.}}]{Abbott:2016blz}%
  \BibitemOpen
  \bibfield  {author} {\bibinfo {author} {\bibfnamefont {B.}~\bibnamefont
  {Abbott}} \emph {et~al.} (\bibinfo {collaboration} {LIGO Scientific,
  Virgo}),\ }\href {\doibase 10.1103/PhysRevLett.116.061102} {\bibfield
  {journal} {\bibinfo  {journal} {Phys. Rev. Lett.}\ }\textbf {\bibinfo
  {volume} {116}},\ \bibinfo {pages} {061102} (\bibinfo {year} {2016})},\
  \Eprint {http://arxiv.org/abs/1602.03837} {arXiv:1602.03837 [gr-qc]}
  \BibitemShut {NoStop}%
\bibitem [{\citenamefont {Psaltis}(2019)}]{Psaltis:2018xkc}%
  \BibitemOpen
  \bibfield  {author} {\bibinfo {author} {\bibfnamefont {D.}~\bibnamefont
  {Psaltis}},\ }\href {\doibase 10.1007/s10714-019-2611-5} {\bibfield
  {journal} {\bibinfo  {journal} {Gen. Rel. Grav.}\ }\textbf {\bibinfo {volume}
  {51}},\ \bibinfo {pages} {137} (\bibinfo {year} {2019})},\ \Eprint
  {http://arxiv.org/abs/1806.09740} {arXiv:1806.09740 [astro-ph.HE]}
  \BibitemShut {NoStop}%
\bibitem [{\citenamefont {Blázquez-Salcedo}\ \emph {et~al.}(2020)\citenamefont
  {Blázquez-Salcedo}, \citenamefont {Herdeiro}, \citenamefont {Kunz},
  \citenamefont {Pombo},\ and\ \citenamefont {Radu}}]{HotColdBald}%
  \BibitemOpen
  \bibfield  {author} {\bibinfo {author} {\bibfnamefont {J.~L.}\ \bibnamefont
  {Blázquez-Salcedo}}, \bibinfo {author} {\bibfnamefont {C.~A.~R.}\
  \bibnamefont {Herdeiro}}, \bibinfo {author} {\bibfnamefont {J.}~\bibnamefont
  {Kunz}}, \bibinfo {author} {\bibfnamefont {A.~M.}\ \bibnamefont {Pombo}}, \
  and\ \bibinfo {author} {\bibfnamefont {E.}~\bibnamefont {Radu}},\ }\href@noop
  {} {\  (\bibinfo {year} {2020})},\ \Eprint {http://arxiv.org/abs/2002.00963}
  {arXiv:2002.00963 [gr-qc]} \BibitemShut {NoStop}%
\bibitem [{\citenamefont {Myung}\ and\ \citenamefont
  {Zou}(2019{\natexlab{a}})}]{Myung:2018vug}%
  \BibitemOpen
  \bibfield  {author} {\bibinfo {author} {\bibfnamefont {Y.~S.}\ \bibnamefont
  {Myung}}\ and\ \bibinfo {author} {\bibfnamefont {D.-C.}\ \bibnamefont
  {Zou}},\ }\href {\doibase 10.1140/epjc/s10052-019-6792-6} {\bibfield
  {journal} {\bibinfo  {journal} {Eur. Phys. J.}\ }\textbf {\bibinfo {volume}
  {C79}},\ \bibinfo {pages} {273} (\bibinfo {year} {2019}{\natexlab{a}})},\
  \Eprint {http://arxiv.org/abs/1808.02609} {arXiv:1808.02609 [gr-qc]}
  \BibitemShut {NoStop}%
\bibitem [{\citenamefont {Boskovic}\ \emph {et~al.}(2019)\citenamefont
  {Boskovic}, \citenamefont {Brito}, \citenamefont {Cardoso}, \citenamefont
  {Ikeda},\ and\ \citenamefont {Witek}}]{Boskovic:2018lkj}%
  \BibitemOpen
  \bibfield  {author} {\bibinfo {author} {\bibfnamefont {M.}~\bibnamefont
  {Boskovic}}, \bibinfo {author} {\bibfnamefont {R.}~\bibnamefont {Brito}},
  \bibinfo {author} {\bibfnamefont {V.}~\bibnamefont {Cardoso}}, \bibinfo
  {author} {\bibfnamefont {T.}~\bibnamefont {Ikeda}}, \ and\ \bibinfo {author}
  {\bibfnamefont {H.}~\bibnamefont {Witek}},\ }\href {\doibase
  10.1103/PhysRevD.99.035006} {\bibfield  {journal} {\bibinfo  {journal} {Phys.
  Rev.}\ }\textbf {\bibinfo {volume} {D99}},\ \bibinfo {pages} {035006}
  (\bibinfo {year} {2019})},\ \Eprint {http://arxiv.org/abs/1811.04945}
  {arXiv:1811.04945 [gr-qc]} \BibitemShut {NoStop}%
\bibitem [{\citenamefont {{Myung}}\ and\ \citenamefont
  {{Zou}}(2018)}]{EMS-Stability}%
  \BibitemOpen
  \bibfield  {author} {\bibinfo {author} {\bibfnamefont {Y.~S.}\ \bibnamefont
  {{Myung}}}\ and\ \bibinfo {author} {\bibfnamefont {D.-C.}\ \bibnamefont
  {{Zou}}},\ }\href@noop {} {\bibfield  {journal} {\bibinfo  {journal} {arXiv
  e-prints}\ ,\ \bibinfo {eid} {arXiv:1812.03604}} (\bibinfo {year} {2018})},\
  \Eprint {http://arxiv.org/abs/1812.03604} {arXiv:1812.03604 [gr-qc]}
  \BibitemShut {NoStop}%
\bibitem [{\citenamefont {Brihaye}\ and\ \citenamefont
  {Hartmann}(2019)}]{Brihaye:2019kvj}%
  \BibitemOpen
  \bibfield  {author} {\bibinfo {author} {\bibfnamefont {Y.}~\bibnamefont
  {Brihaye}}\ and\ \bibinfo {author} {\bibfnamefont {B.}~\bibnamefont
  {Hartmann}},\ }\href {\doibase 10.1016/j.physletb.2019.03.043} {\bibfield
  {journal} {\bibinfo  {journal} {Phys. Lett.}\ }\textbf {\bibinfo {volume}
  {B792}},\ \bibinfo {pages} {244} (\bibinfo {year} {2019})},\ \Eprint
  {http://arxiv.org/abs/1902.05760} {arXiv:1902.05760 [gr-qc]} \BibitemShut
  {NoStop}%
\bibitem [{\citenamefont {Herdeiro}\ and\ \citenamefont
  {Oliveira}(2019)}]{Herdeiro:2019oqp}%
  \BibitemOpen
  \bibfield  {author} {\bibinfo {author} {\bibfnamefont {C.~A.~R.}\
  \bibnamefont {Herdeiro}}\ and\ \bibinfo {author} {\bibfnamefont {J.~M.~S.}\
  \bibnamefont {Oliveira}},\ }\href {\doibase 10.1088/1361-6382/ab1859}
  {\bibfield  {journal} {\bibinfo  {journal} {Class. Quant. Grav.}\ }\textbf
  {\bibinfo {volume} {36}},\ \bibinfo {pages} {105015} (\bibinfo {year}
  {2019})},\ \Eprint {http://arxiv.org/abs/1902.07721} {arXiv:1902.07721
  [gr-qc]} \BibitemShut {NoStop}%
\bibitem [{\citenamefont {Myung}\ and\ \citenamefont
  {Zou}(2019{\natexlab{b}})}]{Myung:2019oua}%
  \BibitemOpen
  \bibfield  {author} {\bibinfo {author} {\bibfnamefont {Y.~S.}\ \bibnamefont
  {Myung}}\ and\ \bibinfo {author} {\bibfnamefont {D.-C.}\ \bibnamefont
  {Zou}},\ }\href {\doibase 10.1140/epjc/s10052-019-7176-7} {\bibfield
  {journal} {\bibinfo  {journal} {Eur. Phys. J.}\ }\textbf {\bibinfo {volume}
  {C79}},\ \bibinfo {pages} {641} (\bibinfo {year} {2019}{\natexlab{b}})},\
  \Eprint {http://arxiv.org/abs/1904.09864} {arXiv:1904.09864 [gr-qc]}
  \BibitemShut {NoStop}%
\bibitem [{\citenamefont {Konoplya}\ and\ \citenamefont
  {Zhidenko}(2019)}]{Konoplya:2019goy}%
  \BibitemOpen
  \bibfield  {author} {\bibinfo {author} {\bibfnamefont {R.~A.}\ \bibnamefont
  {Konoplya}}\ and\ \bibinfo {author} {\bibfnamefont {A.}~\bibnamefont
  {Zhidenko}},\ }\href {\doibase 10.1103/PhysRevD.100.044015} {\bibfield
  {journal} {\bibinfo  {journal} {Phys. Rev.}\ }\textbf {\bibinfo {volume}
  {D100}},\ \bibinfo {pages} {044015} (\bibinfo {year} {2019})},\ \Eprint
  {http://arxiv.org/abs/1907.05551} {arXiv:1907.05551 [gr-qc]} \BibitemShut
  {NoStop}%
\bibitem [{\citenamefont {Herdeiro}\ and\ \citenamefont
  {Oliveira}(2020)}]{Herdeiro:2019tmb}%
  \BibitemOpen
  \bibfield  {author} {\bibinfo {author} {\bibfnamefont {C.~A.~R.}\
  \bibnamefont {Herdeiro}}\ and\ \bibinfo {author} {\bibfnamefont {J.~M.~S.}\
  \bibnamefont {Oliveira}},\ }\href {\doibase 10.1016/j.physletb.2019.135076}
  {\bibfield  {journal} {\bibinfo  {journal} {Phys. Lett.}\ }\textbf {\bibinfo
  {volume} {B800}},\ \bibinfo {pages} {135076} (\bibinfo {year} {2020})},\
  \Eprint {http://arxiv.org/abs/1909.08915} {arXiv:1909.08915 [gr-qc]}
  \BibitemShut {NoStop}%
\bibitem [{\citenamefont {Brihaye}\ \emph {et~al.}(2020)\citenamefont
  {Brihaye}, \citenamefont {Herdeiro},\ and\ \citenamefont
  {Radu}}]{Brihaye:2019gla}%
  \BibitemOpen
  \bibfield  {author} {\bibinfo {author} {\bibfnamefont {Y.}~\bibnamefont
  {Brihaye}}, \bibinfo {author} {\bibfnamefont {C.}~\bibnamefont {Herdeiro}}, \
  and\ \bibinfo {author} {\bibfnamefont {E.}~\bibnamefont {Radu}},\ }\href
  {\doibase 10.1016/j.physletb.2020.135269} {\bibfield  {journal} {\bibinfo
  {journal} {Phys. Lett.}\ }\textbf {\bibinfo {volume} {B802}},\ \bibinfo
  {pages} {135269} (\bibinfo {year} {2020})},\ \Eprint
  {http://arxiv.org/abs/1910.05286} {arXiv:1910.05286 [gr-qc]} \BibitemShut
  {NoStop}%
\bibitem [{\citenamefont {Hod}(2019)}]{Hod:2020ljo}%
  \BibitemOpen
  \bibfield  {author} {\bibinfo {author} {\bibfnamefont {S.}~\bibnamefont
  {Hod}},\ }\href@noop {} {\bibfield  {journal} {\bibinfo  {journal} {Phys.
  Lett.}\ }\textbf {\bibinfo {volume} {B798}},\ \bibinfo {pages} {135025}
  (\bibinfo {year} {2019})},\ \Eprint {http://arxiv.org/abs/2002.01948}
  {arXiv:2002.01948 [gr-qc]} \BibitemShut {NoStop}%
\bibitem [{\citenamefont {Zou}\ and\ \citenamefont
  {Myung}(2019)}]{MassTermMyung}%
  \BibitemOpen
  \bibfield  {author} {\bibinfo {author} {\bibfnamefont {D.-C.}\ \bibnamefont
  {Zou}}\ and\ \bibinfo {author} {\bibfnamefont {Y.~S.}\ \bibnamefont
  {Myung}},\ }\href {\doibase 10.1103/PhysRevD.100.124055} {\bibfield
  {journal} {\bibinfo  {journal} {Phys. Rev.}\ }\textbf {\bibinfo {volume}
  {D100}},\ \bibinfo {pages} {124055} (\bibinfo {year} {2019})},\ \Eprint
  {http://arxiv.org/abs/1909.11859} {arXiv:1909.11859 [gr-qc]} \BibitemShut
  {NoStop}%
\bibitem [{\citenamefont {Herdeiro}\ \emph {et~al.}(2015)\citenamefont
  {Herdeiro}, \citenamefont {Radu},\ and\ \citenamefont
  {Rúnarsson}}]{Herdeiro:2015tia}%
  \BibitemOpen
  \bibfield  {author} {\bibinfo {author} {\bibfnamefont {C.~A.~R.}\
  \bibnamefont {Herdeiro}}, \bibinfo {author} {\bibfnamefont {E.}~\bibnamefont
  {Radu}}, \ and\ \bibinfo {author} {\bibfnamefont {H.}~\bibnamefont
  {Rúnarsson}},\ }\href {\doibase 10.1103/PhysRevD.92.084059} {\bibfield
  {journal} {\bibinfo  {journal} {Phys. Rev.}\ }\textbf {\bibinfo {volume}
  {D92}},\ \bibinfo {pages} {084059} (\bibinfo {year} {2015})},\ \Eprint
  {http://arxiv.org/abs/1509.02923} {arXiv:1509.02923 [gr-qc]} \BibitemShut
  {NoStop}%
\bibitem [{\citenamefont {Herdeiro}\ \emph {et~al.}(2016)\citenamefont
  {Herdeiro}, \citenamefont {Radu},\ and\ \citenamefont
  {Rúnarsson}}]{Herdeiro:2016gxs}%
  \BibitemOpen
  \bibfield  {author} {\bibinfo {author} {\bibfnamefont {C.~A.~R.}\
  \bibnamefont {Herdeiro}}, \bibinfo {author} {\bibfnamefont {E.}~\bibnamefont
  {Radu}}, \ and\ \bibinfo {author} {\bibfnamefont {H.~F.}\ \bibnamefont
  {Rúnarsson}},\ }\bibfield  {booktitle} {\emph {\bibinfo {booktitle}
  {{Proceedings, 3rd Amazonian Symposium on Physics: Belem, Brazil, September
  28-October 2, 2015}}},\ }\href {\doibase 10.1142/S0218271816410145}
  {\bibfield  {journal} {\bibinfo  {journal} {Int. J. Mod. Phys.}\ }\textbf
  {\bibinfo {volume} {D25}},\ \bibinfo {pages} {1641014} (\bibinfo {year}
  {2016})},\ \Eprint {http://arxiv.org/abs/1604.06202} {arXiv:1604.06202
  [gr-qc]} \BibitemShut {NoStop}%
\bibitem [{\citenamefont {Bardeen}\ \emph {et~al.}(1973)\citenamefont
  {Bardeen}, \citenamefont {Carter},\ and\ \citenamefont {Hawking}}]{4lawsBH}%
  \BibitemOpen
  \bibfield  {author} {\bibinfo {author} {\bibfnamefont {J.~M.}\ \bibnamefont
  {Bardeen}}, \bibinfo {author} {\bibfnamefont {B.}~\bibnamefont {Carter}}, \
  and\ \bibinfo {author} {\bibfnamefont {S.~W.}\ \bibnamefont {Hawking}},\
  }\href {https://projecteuclid.org:443/euclid.cmp/1103858973} {\bibfield
  {journal} {\bibinfo  {journal} {Comm. Math. Phys.}\ }\textbf {\bibinfo
  {volume} {31}},\ \bibinfo {pages} {161} (\bibinfo {year} {1973})}\BibitemShut
  {NoStop}%
\bibitem [{\citenamefont {Smarr}(1973)}]{Smarr}%
  \BibitemOpen
  \bibfield  {author} {\bibinfo {author} {\bibfnamefont {L.}~\bibnamefont
  {Smarr}},\ }\href {\doibase 10.1103/PhysRevLett.30.521,
  10.1103/PhysRevLett.30.71} {\bibfield  {journal} {\bibinfo  {journal} {Phys.
  Rev. Lett.}\ }\textbf {\bibinfo {volume} {30}},\ \bibinfo {pages} {71}
  (\bibinfo {year} {1973})},\ \bibinfo {note} {[Erratum: Phys. Rev.
  Lett.30,521(1973)]}\BibitemShut {NoStop}%
\bibitem [{\citenamefont {Derrick}(1964)}]{Derrick:1964ww}%
  \BibitemOpen
  \bibfield  {author} {\bibinfo {author} {\bibfnamefont {G.~H.}\ \bibnamefont
  {Derrick}},\ }\href {\doibase 10.1063/1.1704233} {\bibfield  {journal}
  {\bibinfo  {journal} {J. Math. Phys.}\ }\textbf {\bibinfo {volume} {5}},\
  \bibinfo {pages} {1252} (\bibinfo {year} {1964})}\BibitemShut {NoStop}%
\bibitem [{\citenamefont {{Herdeiro}}\ and\ \citenamefont
  {{Radu}}(2015)}]{ScalarHairReview}%
  \BibitemOpen
  \bibfield  {author} {\bibinfo {author} {\bibfnamefont {C.~A.~R.}\
  \bibnamefont {{Herdeiro}}}\ and\ \bibinfo {author} {\bibfnamefont
  {E.}~\bibnamefont {{Radu}}},\ }\href {\doibase 10.1142/S0218271815420146}
  {\bibfield  {journal} {\bibinfo  {journal} {International Journal of Modern
  Physics D}\ }\textbf {\bibinfo {volume} {24}},\ \bibinfo {eid} {1542014-219}
  (\bibinfo {year} {2015})},\ \Eprint {http://arxiv.org/abs/1504.08209}
  {arXiv:1504.08209 [gr-qc]} \BibitemShut {NoStop}%
\bibitem [{\citenamefont {Gibbons}\ and\ \citenamefont
  {Wiltshire}(1986)}]{DilatonGibbons2}%
  \BibitemOpen
  \bibfield  {author} {\bibinfo {author} {\bibfnamefont {G.~W.}\ \bibnamefont
  {Gibbons}}\ and\ \bibinfo {author} {\bibfnamefont {D.~L.}\ \bibnamefont
  {Wiltshire}},\ }\href {\doibase 10.1016/S0003-4916(86)80012-4,
  10.1016/0003-4916(87)90008-X} {\bibfield  {journal} {\bibinfo  {journal}
  {Annals Phys.}\ }\textbf {\bibinfo {volume} {167}},\ \bibinfo {pages} {201}
  (\bibinfo {year} {1986})},\ \bibinfo {note} {[Erratum: Annals
  Phys.176,393(1987)]}\BibitemShut {NoStop}%
\bibitem [{\citenamefont {Gregory}\ and\ \citenamefont
  {Harvey}(1993)}]{Gregory:1992kr}%
  \BibitemOpen
  \bibfield  {author} {\bibinfo {author} {\bibfnamefont {R.}~\bibnamefont
  {Gregory}}\ and\ \bibinfo {author} {\bibfnamefont {J.~A.}\ \bibnamefont
  {Harvey}},\ }\href {\doibase 10.1103/PhysRevD.47.2411} {\bibfield  {journal}
  {\bibinfo  {journal} {Phys. Rev.}\ }\textbf {\bibinfo {volume} {D47}},\
  \bibinfo {pages} {2411} (\bibinfo {year} {1993})},\ \Eprint
  {http://arxiv.org/abs/hep-th/9209070} {arXiv:hep-th/9209070 [hep-th]}
  \BibitemShut {NoStop}%
\bibitem [{\citenamefont {Horne}\ and\ \citenamefont
  {Horowitz}(1993)}]{Horne:1992bi}%
  \BibitemOpen
  \bibfield  {author} {\bibinfo {author} {\bibfnamefont {J.~H.}\ \bibnamefont
  {Horne}}\ and\ \bibinfo {author} {\bibfnamefont {G.~T.}\ \bibnamefont
  {Horowitz}},\ }\href {\doibase 10.1016/0550-3213(93)90621-U} {\bibfield
  {journal} {\bibinfo  {journal} {Nucl. Phys.}\ }\textbf {\bibinfo {volume}
  {B399}},\ \bibinfo {pages} {169} (\bibinfo {year} {1993})},\ \Eprint
  {http://arxiv.org/abs/hep-th/9210012} {arXiv:hep-th/9210012 [hep-th]}
  \BibitemShut {NoStop}%
\bibitem [{\citenamefont {{Berti}}\ \emph {et~al.}(2009)\citenamefont
  {{Berti}}, \citenamefont {{Cardoso}},\ and\ \citenamefont
  {{Starinets}}}]{qnmcardoso}%
  \BibitemOpen
  \bibfield  {author} {\bibinfo {author} {\bibfnamefont {E.}~\bibnamefont
  {{Berti}}}, \bibinfo {author} {\bibfnamefont {V.}~\bibnamefont {{Cardoso}}},
  \ and\ \bibinfo {author} {\bibfnamefont {A.~O.}\ \bibnamefont
  {{Starinets}}},\ }\href {\doibase 10.1088/0264-9381/26/16/163001} {\bibfield
  {journal} {\bibinfo  {journal} {Classical and Quantum Gravity}\ }\textbf
  {\bibinfo {volume} {26}},\ \bibinfo {eid} {163001} (\bibinfo {year}
  {2009})},\ \Eprint {http://arxiv.org/abs/0905.2975} {arXiv:0905.2975 [gr-qc]}
  \BibitemShut {NoStop}%
\bibitem [{\citenamefont {Messiah}(1961)}]{Messiah:1961}%
  \BibitemOpen
  \bibfield  {author} {\bibinfo {author} {\bibfnamefont {A.}~\bibnamefont
  {Messiah}},\ }\href@noop {} {\emph {\bibinfo {title} {{QUANTUM MECHANICS,
  Chapter III2}}}}\ (\bibinfo  {publisher} {North Holland Publishing Company},\
  \bibinfo {year} {1961})\BibitemShut {NoStop}%
\bibitem [{\citenamefont {Kimura}(2017)}]{Kimura:2017uor}%
  \BibitemOpen
  \bibfield  {author} {\bibinfo {author} {\bibfnamefont {M.}~\bibnamefont
  {Kimura}},\ }\href {\doibase 10.1088/1361-6382/aa903f} {\bibfield  {journal}
  {\bibinfo  {journal} {Class. Quant. Grav.}\ }\textbf {\bibinfo {volume}
  {34}},\ \bibinfo {pages} {235007} (\bibinfo {year} {2017})},\ \Eprint
  {http://arxiv.org/abs/1706.01447} {arXiv:1706.01447 [gr-qc]} \BibitemShut
  {NoStop}%
\bibitem [{\citenamefont {Kimura}\ and\ \citenamefont
  {Tanaka}(2018)}]{Kimura:2018eiv}%
  \BibitemOpen
  \bibfield  {author} {\bibinfo {author} {\bibfnamefont {M.}~\bibnamefont
  {Kimura}}\ and\ \bibinfo {author} {\bibfnamefont {T.}~\bibnamefont
  {Tanaka}},\ }\href {\doibase 10.1088/1361-6382/aadc13} {\bibfield  {journal}
  {\bibinfo  {journal} {Class. Quant. Grav.}\ }\textbf {\bibinfo {volume}
  {35}},\ \bibinfo {pages} {195008} (\bibinfo {year} {2018})},\ \Eprint
  {http://arxiv.org/abs/1805.08625} {arXiv:1805.08625 [gr-qc]} \BibitemShut
  {NoStop}%
\bibitem [{\citenamefont {Kimura}\ and\ \citenamefont
  {Tanaka}(2019)}]{Kimura:2018whv}%
  \BibitemOpen
  \bibfield  {author} {\bibinfo {author} {\bibfnamefont {M.}~\bibnamefont
  {Kimura}}\ and\ \bibinfo {author} {\bibfnamefont {T.}~\bibnamefont
  {Tanaka}},\ }\href {\doibase 10.1088/1361-6382/ab0193} {\bibfield  {journal}
  {\bibinfo  {journal} {Class. Quant. Grav.}\ }\textbf {\bibinfo {volume}
  {36}},\ \bibinfo {pages} {055005} (\bibinfo {year} {2019})},\ \Eprint
  {http://arxiv.org/abs/1809.00795} {arXiv:1809.00795 [gr-qc]} \BibitemShut
  {NoStop}%
\bibitem [{\citenamefont {Berti}\ \emph {et~al.}(2015)\citenamefont {Berti}
  \emph {et~al.}}]{Berti:2015itd}%
  \BibitemOpen
  \bibfield  {author} {\bibinfo {author} {\bibfnamefont {E.}~\bibnamefont
  {Berti}} \emph {et~al.},\ }\href {\doibase 10.1088/0264-9381/32/24/243001}
  {\bibfield  {journal} {\bibinfo  {journal} {Class. Quant. Grav.}\ }\textbf
  {\bibinfo {volume} {32}},\ \bibinfo {pages} {243001} (\bibinfo {year}
  {2015})},\ \Eprint {http://arxiv.org/abs/1501.07274} {arXiv:1501.07274
  [gr-qc]} \BibitemShut {NoStop}%
\bibitem [{\citenamefont {Yagi}\ \emph {et~al.}(2012)\citenamefont {Yagi},
  \citenamefont {Stein}, \citenamefont {Yunes},\ and\ \citenamefont
  {Tanaka}}]{Yagi:2011xp}%
  \BibitemOpen
  \bibfield  {author} {\bibinfo {author} {\bibfnamefont {K.}~\bibnamefont
  {Yagi}}, \bibinfo {author} {\bibfnamefont {L.~C.}\ \bibnamefont {Stein}},
  \bibinfo {author} {\bibfnamefont {N.}~\bibnamefont {Yunes}}, \ and\ \bibinfo
  {author} {\bibfnamefont {T.}~\bibnamefont {Tanaka}},\ }\href {\doibase
  10.1103/PhysRevD.85.064022} {\bibfield  {journal} {\bibinfo  {journal} {Phys.
  Rev. D}\ }\textbf {\bibinfo {volume} {85}},\ \bibinfo {pages} {064022}
  (\bibinfo {year} {2012})},\ \bibinfo {note} {[Erratum: Phys.Rev.D 93, 029902
  (2016)]},\ \Eprint {http://arxiv.org/abs/1110.5950} {arXiv:1110.5950 [gr-qc]}
  \BibitemShut {NoStop}%
\bibitem [{\citenamefont {Berti}\ \emph {et~al.}(2018)\citenamefont {Berti},
  \citenamefont {Yagi},\ and\ \citenamefont {Yunes}}]{Berti:2018cxi}%
  \BibitemOpen
  \bibfield  {author} {\bibinfo {author} {\bibfnamefont {E.}~\bibnamefont
  {Berti}}, \bibinfo {author} {\bibfnamefont {K.}~\bibnamefont {Yagi}}, \ and\
  \bibinfo {author} {\bibfnamefont {N.}~\bibnamefont {Yunes}},\ }\href
  {\doibase 10.1007/s10714-018-2362-8} {\bibfield  {journal} {\bibinfo
  {journal} {Gen. Rel. Grav.}\ }\textbf {\bibinfo {volume} {50}},\ \bibinfo
  {pages} {46} (\bibinfo {year} {2018})},\ \Eprint
  {http://arxiv.org/abs/1801.03208} {arXiv:1801.03208 [gr-qc]} \BibitemShut
  {NoStop}%
\end{thebibliography}%

\end{document}